\documentclass{emulateapj}
\usepackage{color}
\usepackage{amsmath}
\usepackage{helvet}
\usepackage{mathtools}
\usepackage[caption=false]{subfig}
\begin{document}

\title{SELECTION OF BURST-LIKE TRANSIENTS AND STOCHASTIC VARIABLES USING MULTI-BAND IMAGE DIFFERENCING IN THE PAN-STARRS1 MEDIUM-DEEP SURVEY}

\author{S. Kumar\altaffilmark{1}, S. Gezari\altaffilmark{1}, S. Heinis\altaffilmark{1}, R. Chornock\altaffilmark{2}, E. Berger\altaffilmark{2}, A. Rest\altaffilmark{3}, M.E. Huber\altaffilmark{4}, R.J. Foley\altaffilmark{6,10}, G. Narayan\altaffilmark{4}, G.H. Marion\altaffilmark{4}, D. Scolnic\altaffilmark{7}, A. Soderberg\altaffilmark{2}, A. Lawrence\altaffilmark{5}, C.W. Stubbs\altaffilmark{2}, R.P. Kirshner\altaffilmark{2}, A.G. Riess\altaffilmark{7}, S.J. Smartt\altaffilmark{8}, K. Smith\altaffilmark{8}, W.M Wood-Vasey\altaffilmark{9}, 
W. S. Burgett\altaffilmark{4}, 
K. C. Chambers\altaffilmark{4}, 
H. Flewelling\altaffilmark{4},
N. Kaiser\altaffilmark{4}, 
N. Metcalfe\altaffilmark{3}, 
P. A. Price\altaffilmark{4}, 
J. L. Tonry\altaffilmark{4}, 
\& R. J. Wainscoat\altaffilmark{4}}

\altaffiltext{1}{Department of Astronomy, University of Maryland, Stadium Drive, College Park, MD  21224, USA}
\altaffiltext{2}{Harvard-Smithsonian Center for Astrophysics, 60 Garden Street, Cambridge, Massachusetts 02138, USA}
\altaffiltext{3}{Space Telescope Science Institute, 3700 San Martin Drive, Baltimore, Maryland 21218, USA}
\altaffiltext{4}{Institute for Astronomy, University of Hawaii, 2680 Woodlawn Drive, Honolulu, Hawaii 96822, USA}
\altaffiltext{5}{Institute for Astronomy, University of Edinburgh Scottish Universities Physics Alliance, Royal Observatory, Blackford Hill, Edinburgh EH9 3HJ, UK}
\altaffiltext{6}{Astronomy Department, University of Illinois at Urbana-Champaign, 1002 West Green Street, Urbana, IL 61801, USA}
\altaffiltext{7}{Department of Physics and Astronomy, Johns Hopkins University, 3400 North Charles Street, Baltimore, Maryland 21218, USA}
\altaffiltext{8}{Astrophysics Research Centre, School of Mathematics and Physics, Queen's University Belfast, Belfast BT7 1NN, UK}
\altaffiltext{9}{Pittsburgh Particle Physics, Astrophysics, and Cosmology Center, Department of Physics and Astronomy, University of Pittsburgh, 3941 O'Hara Street, Pittsburgh, Pennsylvania 15260, USA}
\altaffiltext{10}{Department of Physics,University of Illinois Urbana-Champaign,1110 W.\ Green Street, Urbana, IL 61801 USA}

\begin{abstract}

We present a novel method for the light-curve characterization of Pan-STARRS1 Medium Deep Survey (PS1 MDS) extragalactic sources into stochastic variables (SV) and burst-like (BL) transients, using multi-band image-differencing time-series data. We select detections in difference images associated with galaxy hosts using a star/galaxy catalog extracted from the deep PS1 MDS stacked images, and adopt a maximum a posteriori formulation to model their difference-flux time-series in four Pan-STARRS1 photometric bands $g_{P1}, r_{P1}, i_{P1}$, and $z_{P1}$. We use three deterministic light-curve models to fit burst-like transients; a Gaussian, a Gamma distribution, and an analytic supernova (SN) model, and one stochastic light curve model, the Ornstein-Uhlenbeck process, in order to fit variability that is characteristic of active galactic nuclei (AGN). We assess the quality of fit of the models band-wise source-wise, using their estimated leave-out-one cross-validation likelihoods and corrected Akaike information criteria. We then apply a K-means clustering algorithm on these statistics, to determine the source classification in each band. The final source classification is derived as a combination of the individual filter classifications, resulting in two measures of classification quality, from the averages across the photometric filters of 1) the classifications determined from the closest K-means cluster centers, and 2) the square distances from the clustering centers in the K-means clustering spaces. For a verification set of active galactic nuclei and supernovae, we show that SV and BL occupy distinct regions in the plane constituted by these measures. We use our clustering method to characterize $4361$ extragalactic image difference detected sources in the first 2.5 years of the PS1 MDS, into $1529$ BL, and $2262$ SV, with a purity of $95.00\%$ for AGN, and $90.97\%$ for SN based on our verification sets. We combine our light-curve classifications with their nuclear or off-nuclear host galaxy offsets, to define a robust photometric sample of $1233$ active galactic nuclei and $812$ supernovae. With these two samples, we characterize their variability and host galaxy properties, and identify simple photometric priors that would enable their real-time identification in future wide-field synoptic surveys. 

\end{abstract}


\section{Introduction}

With the advent of the LSST era, human-intervened classification (with the exception of citizen science) will become untenable, requiring automated source identification in large volumes of data in archival catalogs, as well as in real-time data. Recent increases in computational resource availability and efficiency have enabled near-complete automated transient discovery in large surveys \citep{Bloom2012}. Machine-learning methods are slowly replacing human judgement for transient classification in real-time, as well as in large survey catalogs \citep{Richards2011,Sanders2014,Pichara2013}. The knowledge of prior event types makes it possible to look for specific events in the data with a high degree of completeness and efficiency using time-variability \citep{Bailer-Jones2012,Butler2010,Schmidt2010,Choi2013}, color based selection \citep{Richards2009}, multi-wavelength catalog associations \citep{Mendez2013}, and host-galaxy properties \citep{Foley2013}. Also, generalized automated machine classification algorithms based on random-forest methods \citep{Kitty2014}, support vector machines and naive Bayes estimates \citep{Mahabal}, and sparse matrix methods \citep{Wozniak2013} that use a number of photometric and non-photometric features have been demonstrated to achieve classifications with very high purity.

As the number of detected transients grows very large in wide-field time domain surveys, complete spectroscopic follow up becomes impossible due to limited resources and faint magnitude limits.  Classification methods using time-series data alone are favorable, and have been applied in the past to a broad range of sources; \citet{Choi2013} discuss the identification of AGN via damped-random walk parameterization of image-differencing light curves, \citet{KimBailer-Jones2013} on the applicability of single and multiple Ornstein-Uhlenbeck (OU) processes  to AGN light curves, and \citet{Butler2010} on the separation of AGN from variable stars in photometric surveys through damped-random walk parameterization. For supernovae (SNe), \citet{Kessler2010} discusses various photometric template fitting methods that enable their identification with particular SN classes. 

\cite{Bailer-Jones2012} gives a powerful method to pick suitable models for deterministic or stochastic lightcurves using leave-one-out cross-validation. This method can be used to determine source type, based on the likelihoods of the various models for a given lightcurve.  So far, the applicability of selecting sources based on time-series modeling has been limited to single-band detections \citep{Choi2013}, or have typically used magnitude time-series data \citep{Butler2010}, which are undefined for negative difference-fluxes. Also, computational limitations typically have lead to the use of only single models as predictors for class, or only using simple statistical criteria for model assessment, rendering classification schemes prone to the possibility of systematic misclassification. 

We attempt here to use time-series methods alone to classify sources into general categories of burst-like (BL) or stochastically variable (SV).  These light-curve classes capture the variability behavior of the two most common extragalactic sources detected in image-differencing surveys, AGN and SNe.  We present a novel method that separates BL and SV sources with high purity, using supervised machine-learning methods. Using multi-band difference-flux in the $g_{\rm P1},r_{\rm P1},i_{\rm P1}$ and $z_{\rm P1}$ bands, we select BL and SV from $4361$ difference-image sources with galaxy hosts. For lightcurves in each band, we estimate the fitnesses of analytical BL models relative to the OU process, using both their estimated leave-out-one cross-validation likelihoods (LOOCV), and corrected-Akaike information criteria (AICc). We show that the use of simple analytical models with suitably chosen priors, which mimic the approximate shapes of BL light curves (predominantly SNe), is sufficient for segregating them from SV, thereby obviating the need for exact models dependent on specific BL subclasses. The model statistical characterizations are combined across sources using a K-means clustering algorithm \citep{Kanungo}, to provide robust source classifications in each filter. The filter-wise classifications are then averaged to give final source classifications. Based on our BL and SV photometric classifications and host galaxy offset cuts, we define a photometric sample of SNe and AGN, which we then use to define observational priors for future surveys such as LSST. 

This paper is organized as follows: In \S\ref{PS1survey} we discuss the details of the survey and our data pipeline; in \S\ref{TS} we discuss time-series models used to describe BL and SV difference-flux light curves; in \S\ref{FITNESS} we elaborate on computing LOOCV and AICc to estimate model fitness; in \S\ref{Classification} we discuss our classification method, characterize the properties of extragalactic variables and transients, and combine our light-curve classifications with host galaxy offsets in order to define a robust photometric sample of SNe and AGN; and in \S\ref{Hosts} we describe the source and host galaxy properties of our variability/offset selected SN and AGN samples, and report on priors that can aid in their real-time classification in future surveys.
 
\begin{deluxetable}{ccc}
\tablecaption{Pan-STARRS1 Medium Deep Survey Field Centers \label{fields}}
\tablenum{1}
\tablehead{\colhead{\nodata} & \colhead{J2000} & \colhead{J2000} \\
\colhead{Field} & \colhead{RA} & \colhead{Declination}\\ 
\colhead{\nodata} & \colhead{Degrees} & \colhead{Degrees}} 
\startdata
\hline
\rule{0pt}{2ex} MD01 & $02^{\rm h}23^{\rm m}30^{\rm s}$ & $-04\deg15\arcmin$ \\
MD02 & $03^{\rm h}32^{\rm m}24^{\rm s}$ & $-27\deg48\arcmin$ \\
MD03 & $08^{\rm h}42^{\rm m}22^{\rm s}$ & $44\deg19\arcmin$ \\
MD04 & $10^{\rm h}00^{\rm m}00^{\rm s}$ & $02\deg12\arcmin$ \\
MD05 & $10^{\rm h}47^{\rm m}40^{\rm s}$ & $58\deg04\arcmin$ \\
MD06 & $12^{\rm h}20^{\rm m}30^{\rm s}$ & $47\deg07\arcmin$ \\
MD07 & $14^{\rm h}14^{\rm m}48^{\rm s}$ & $53\deg04\arcmin$ \\
MD08 & $16^{\rm h}11^{\rm m}08^{\rm s}$ & $54\deg57\arcmin$ \\
MD09 & $22^{\rm h}16^{\rm m}45^{\rm s}$ & $00\deg16\arcmin$ \\
MD10 & $23^{\rm h}29^{\rm m}14^{\rm s}$ & $00\deg25\arcmin$ 
\enddata
\end{deluxetable}

\section{The Pan-STARRS1 Survey and Transient Alert Database} \label{PS1survey}

\begin{figure}[htp]
\includegraphics[height=3in,width=3in]{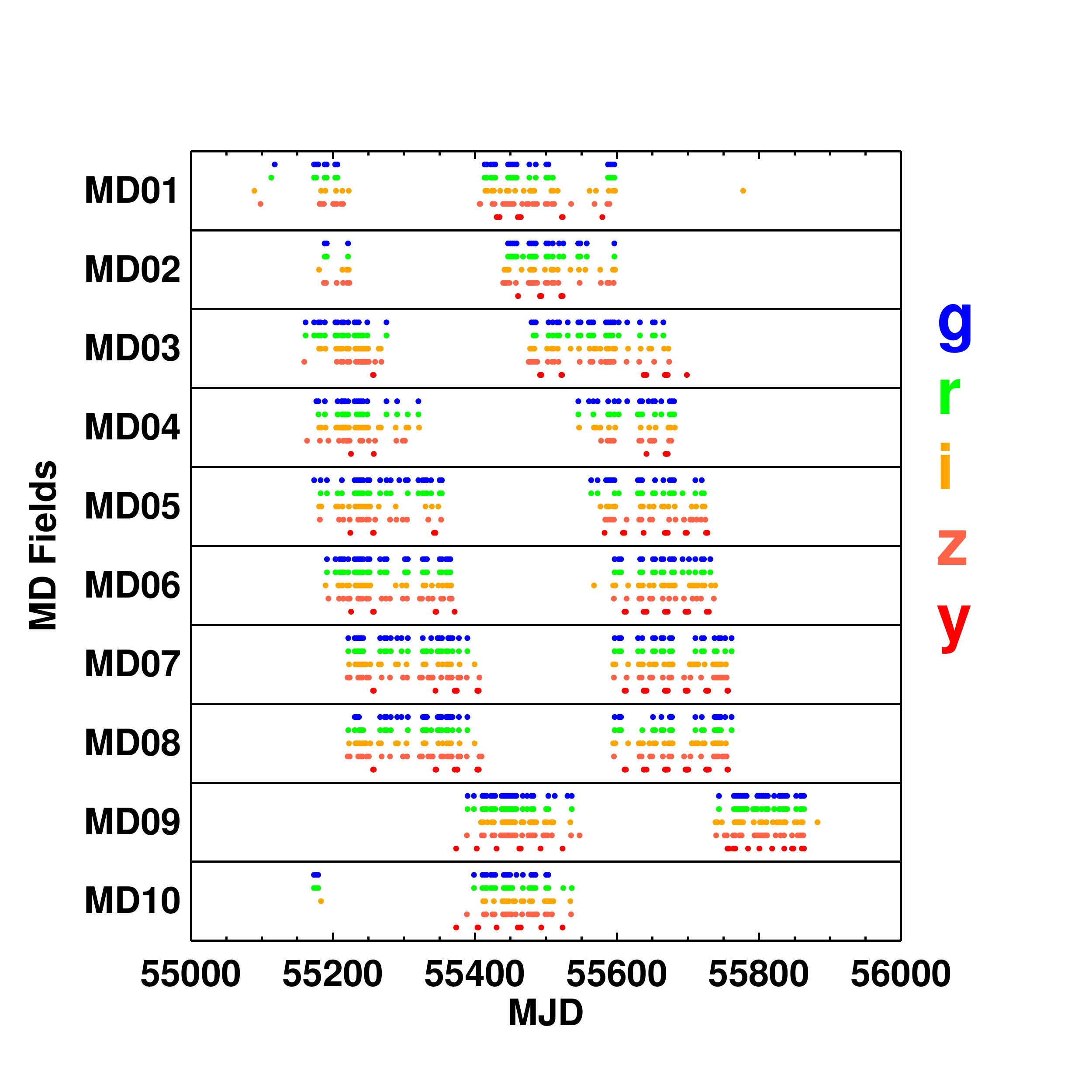}
\centering \caption{The Pan-STARRS1 survey has a staggered 3-day cadence in the $g_{P1},r_{P1},i_{P1}$, and $z_{P1}$ bands corresponding to $6$ observations per month per filter, while $y_{P1}$ is observed during bright-time. The observations we use in this paper extend from $2009$ September $14$ till  $2011$ November $17$. In this paper $y_{P1}$ is not used due to the relatively sparse cadence as compared to the other filters.} \label{Cadence} 
\end{figure}

The Pan-STARRS1 (PS1) telescope \citep{Hodapp2004} is a $1.8$ meter diameter telescope on the summit of Haleakala, Hawaii with a $f /4.4$ primary mirror, and a 0.9 m secondary, delivering an image with a diameter of $3.3$ degrees onto $60$, $4800 \times 4800$ pixel detectors, with $10\mu$m pixels that subtend $0.258"$ each \citep{Tonryonaka2009,Hodapp2004}. The observations are obtained through a set of 5 broadband filters $g_{\rm P1}$,$r_{\rm P1}$,$i_{\rm P1}$,$z_{\rm P1}$,$y_{\rm P1}$, each with a limiting magnitude per nightly epoch of $\sim23.5$ mag. Although the filter system for PS1 has much in common with that used in previous surveys, such as the SDSS, there are substantial differences. For more technical details refer to \citet{Stubbs2010}, and \citet{Tonry2012}. 

The PS1 survey has two operating modes, 1) the $3\pi$ survey which covers $3\pi$ square degrees at $\delta > -30$ degrees in $5$ bands with a cadence of $2$ observations per filter in a $6$ month period, and 2) the Medium Deep Survey (MDS) which obtains deeper multi-epoch images in 5 bands of 10 fields, each $8$ square degrees, listed in Table \ref{fields}, designed for both extensive temporal coverage, and full-survey stacked static-sky depth. Depending on the weather, the accessible fields are observed with a staggered 3-day cadence in each band during dark and gray time ($g_{\rm P1} , r_{\rm P1}$ on the first day, $i_{\rm P1}$ on the second day, $z_{\rm P1}$ on the third day, and then repeat with $g_{\rm P1} , r_{\rm P1}$), and in the $y_{\rm P1}$ band during bright time. On average, the cadence (Fig.~\ref{Cadence}) is $6$ observations per filter per month, with a $1$ week gap during bright time, during which time the Medium Deep fields are observed exclusively in $y_{\rm P1}$. We require the dense time-series (cadence$\approx$few days) for robust variability-based classifications, thereby making the MD survey our survey of choice. 

\begin{figure*}[htp]
\includegraphics[scale=0.25]{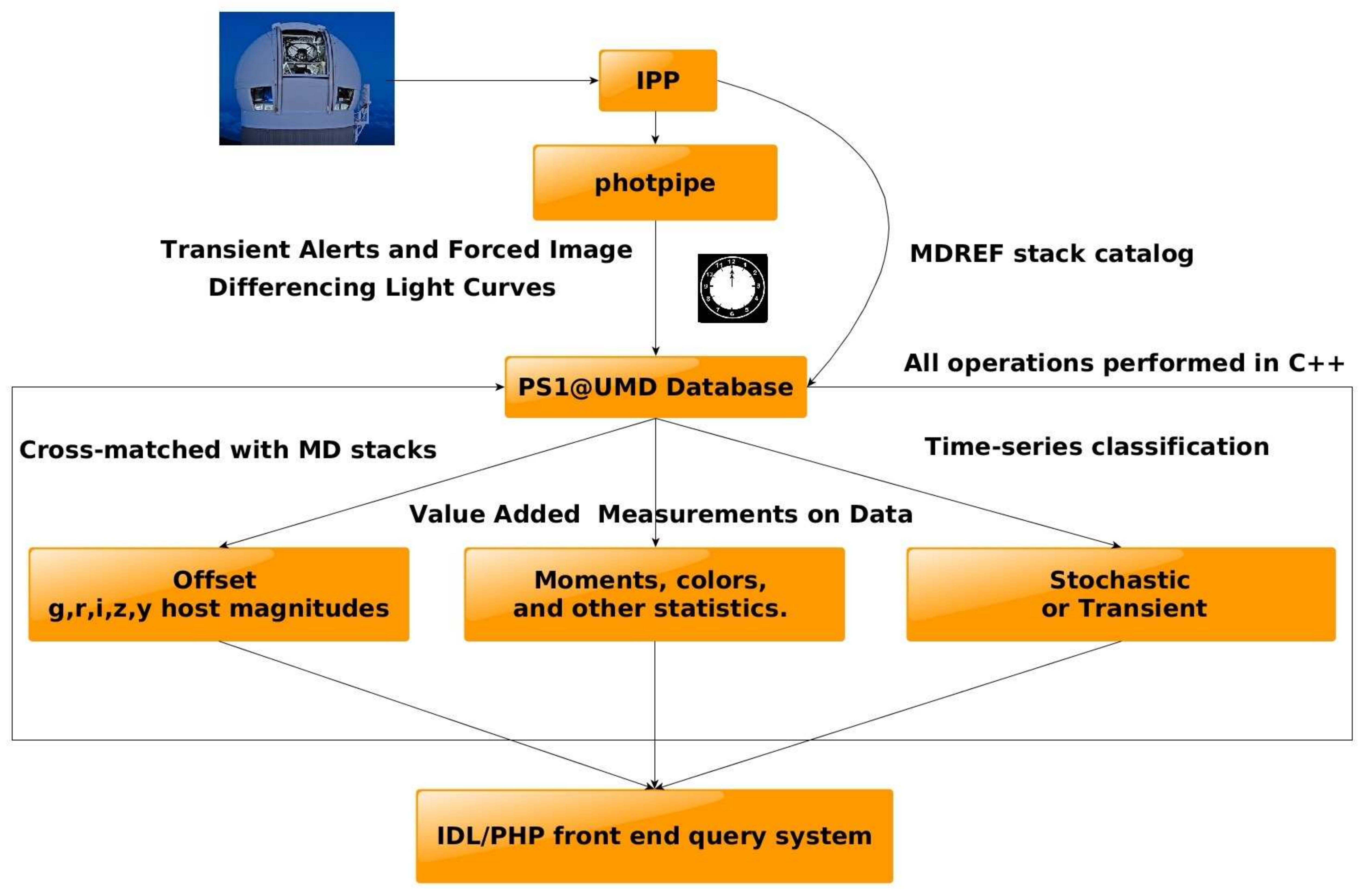}
\centering \caption{The PS1-UMD data pipeline. The data are relayed from the IPP via the \it{photpipe} pipeline that provides transient alerts, as well as performs forced photometry and image differencing. At UMD the data are downloaded, enhanced with statistical parameterizations, and assimilated into SQL databases using a C++ framework, which are then queried using IDL or PHP for interactive web-based analysis.} \label{panstarrseps} 
\end{figure*}

The PS1 MD data are processed using the image processing pipeline (IPP) located in Hawaii. The IPP performs flat-fielding and detrending on each of the individual images using white light flat-field images from a dome screen, in combination with an illumination correction obtained by rastering sources across the field of view. Bad pixel masks are applied, and carried forward for use in the stacking stage. After determining an initial astrometric solution \citep{Magnier08}, the flat-fielded images are then warped onto the tangent plane of the sky, using a flux conserving algorithm. The image scale of the warped images is 0.250 arcsec/pixel. In the MD fields, all images from a given night are collected with eight dithers. This allows the removal of defects like cosmic rays or satellite streaks, before they are combined into a nightly stack using a variance-weighted scheme. Nightly stacks of images, each with a $8$ square degree field of view, as well as seasonal deep stack reference images are created, which are then transferred to the Harvard Faculty of Arts and Sciences Odyssey Research Computing cluster, where they are processed through a frame subtraction analysis using the {\it photpipe} image differencing pipeline originally developed for the SuperMACHO and ESSENCE surveys \citep{Rest2005,Garg2007}. Significant flux excursions are then detected in the difference images using a modified version of the DoPhot\citep{Schechter93,Rest2013} photometry package, and they are tagged as a source, if they satisfy the following conditions: 
  
\begin{itemize}
\item Positive detections with a signal-to-noise ratio
(SNR) $\geq 5$ in at least three images within a time window of $15$ days.
\item Detections in at least two filters.
\item No previous alert at that position.
\end{itemize}

While for source detection, DoPhot requires at least 3 consecutive positive detections within a 15 day period, once the alert is registered, forced photometry can detect both positive and negative fluxes. These criteria remove the majority of ``bogus'' detections due to non-astrophysical sources, such as camera defects, cosmic rays, and image-differencing artifacts.  The PS1 alerts are published to an online alerts database located in Harvard. Our automated pipeline then downloads the alerts database to our local database servers at University of Maryland on a nightly basis. The alerts are then processed and additional value added measurements are made on the data to enable easy characterization of sources via a SQL-IDL-C++ pipeline (Fig.~\ref{panstarrseps}). The sources are automatically cross-matched with custom multi-band deep-stack catalogs \citep{Heinis2014} to derive host associations. Other statistics such as color evolution, and higher moments of magnitude and flux are also computed and stored in our database. Web pages that derive custom cuts on the data based on host properties, host offsets, color, magnitude, and time variability properties are also updated nightly. Our custom query page can be used to query the database and display column-wise sortable results on a web page. The page can also be used to visualize the data in our database using simple $2$ dimensional plots or histograms that are created in IDL which are displayed on a web page. Finally, the transient alerts are classified based on their light curves using our time-series method discussed in \S\ref{Classification}. 

\subsection{Pre-Processing the Alerts for Classification} 

The two most commonly occurring extragalactic time-varying sources are SNe and AGN, which fall under the broad categories of burst-like and stochastically varying, respectively. However, these broad classifications, in combination with host galaxy offsets, also enable us to discover more rare and exotic variables and transients.  For example, nuclear BL sources should include tidal disruption events \citep{Gezari2012, Chornock2014}, off-nuclear BL sources may contain gamma-ray burst afterglows \citep{Cenko2013, Singer2013}, and off-nuclear SV sources may be offset AGN from a post-merger recoiling supermassive black holes \citep{Blecha2011}. 

We identify extragalactic alerts by cross-matching the $18,058$ alerts detected in the first $2.5$ years of the PS1 MDS with galaxies detected in our custom multi-band deep-stack star/galaxy catalogs \citep{Heinis2014}. Galaxies are detected in $\chi^2$ images \citep{Szalay1999} built from CFHT's u-band and the five PS1 bands. The detection threshold, defined by the $\chi^2$ distribution, is equivalent to a SNR of $1.9\sigma$. The photometry is then performed using SExtractor \citep{Bertin1996} in Kron elliptical apertures which are also used in cross matching alerts with the objects in the catalog. The catalog contains $\approx10^7$ objects which have been classified as stars or galaxies with over $90\%$ accuracy for sources with magnitudes $<24$ mag, using an optimized SVM classification scheme that takes into account the shape, color, and magnitudes of the detections \citep{Heinis2014}.  Thus we only select alerts with $i_{\rm P1-host}<24$ mag, where the star/galaxy classification is reliable.   We identified $4361$ extragalactic alerts using the catalog, which we then characterized as SV or BL using multi-band difference-flux time-series.  Note that we do not include extragalactic alerts with unresolved hosts in our sample, such as quasars, or ``hostless'' alerts, those with either a faint host galaxy ($i_{\rm P1-host}>24$ mag), or located outside the elliptical region that defines their host galaxy.

To characterize the sources, we model their difference-flux time-series obtained from forced photometry in each of the $g_{\rm P1},r_{\rm P1},i_{\rm P1}$, and $z_{\rm P1}$ bands, and then combine the characterizations across the filters. Forced photometry, done by the {\it photpipe} pipeline, is obtained by performing PSF fitting photometry on all difference images in each band, at the location of any transient candidate, in order to fully exploit all available difference-flux time-series information on the alert, using data from prior to the alert detection. In our method, we decided to use difference-fluxes instead of differential magnitudes because stochastic light curves can have negative difference-fluxes (if they were brighter in the reference image), for which AB magnitudes cannot be defined.

Before we perform the classification on the difference-flux light curves, we pre-process them to remove artifacts, as well as to make them conform with SV and BL model priors.  Many difference-flux light curves also contain singular large difference flux outliers which affect model classification significantly. To remove them, each difference-flux point $y_i$ in a given light curve is compared to its previous and next difference-flux values, $y_{i-1}$ and $y_{i+1}$, and their photometric errors $dy_{i-1}$ and $dy_{i+1}$,with the criterion that at least one of  $ |y_{i}-y_{i-1}| < 10 dy_{i-1}$, or $|y_{i}-y_{i+1}| < 10 dy_{i+1}$ is satisfied for the difference-flux point to be accepted as non erroneous. Since most difference-flux errors are much larger than this cutoff and most differences in successive difference-flux values are much smaller than this, we can ensure that the light curve is unaffected, while the outliers are removed. Since the starting and ending points of light curves cannot be subject to one of these criteria, we discard them after removing the erroneous difference flux points. Also, our condition will not weed out the turn-on of BL lightcurves, because only $|y_{i}-y_{i-1}| < 10 dy_{i-1}$ may not be satisfied due to a short rise-time, however, $|y_{i}-y_{i+1}| < 10 dy_{i+1}$ will be satisfied for all BL lightcurves, given the Pan-STARRS1 medium-deep survey observations repeat once every 3 days, in the $g_{P1},r_{P1},i_{P1},z_{P1}$ bands. Finally, we also transform the light curves such that the minimum difference-flux is $0$. This is done so as to make them conform to the limits for the priors for BL light curves, especially for light curves where the BL source was active in the reference image. 

\begin{figure*}[htp]
\centering
\includegraphics[scale=0.5]{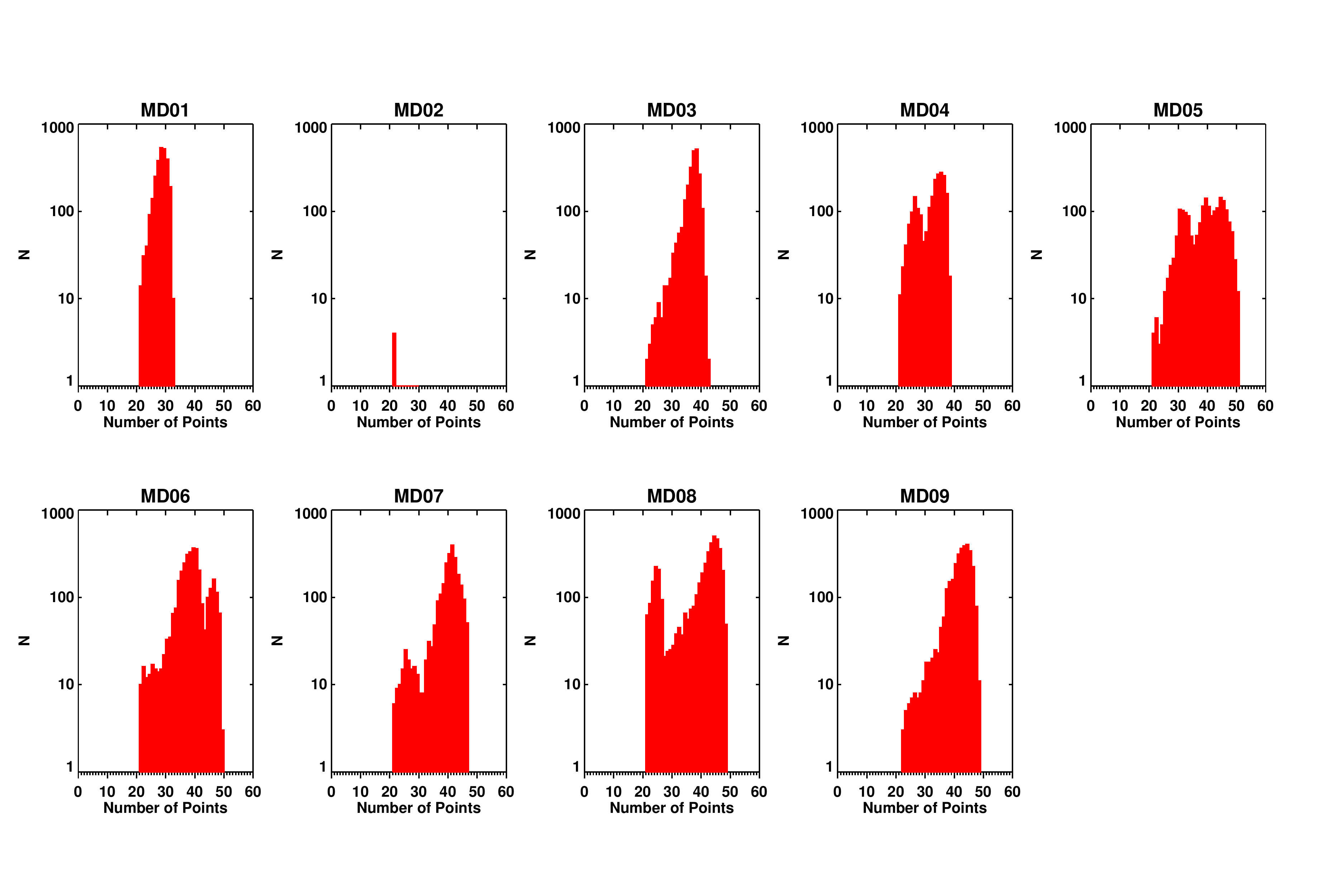}
\caption{Sources plotted in the figure satisfy our criteria for classification: a light curve with $n\geq20$ points in all $4$ filters. MD10 has no points which satisfy our criteria due to only a single full season of coverage in the first two and a half years of the PS1 MDS. } \label{histNPOINTS} 
\end{figure*}

In our analysis we only use light curves which have at least $n=20$ distinct difference-flux measurements post-processing in any filter, so as to ensure that this is at least four times as large as the maximum number of parameters $k_{max}$ used in any of the models (Maximum number of parameters in any time-series model (\S\ref{TS}) is $5$). This is done to prevent over-fitting of the data, which may result in model comparisons not being meaningful. Also, $n=20$ is not a restrictive limit for classification purposes since this is a factor $\approx2$ smaller than the average number of photometric measurements in any filter for all $4361$ sources, which is $\approx 36$. Fig. \ref{histNPOINTS} shows the histogram of number of distinct points in the $g_{\rm P1},r_{\rm P1},i_{\rm P1},$ or $z_{\rm P1}$ filters for all the PS1 transient alerts associated with galaxy hosts that pass our cuts.  In the next section we discuss the time-series models that we use to classify the difference-flux light curves. 

\section{Time-Series Models} \label{TS}

Since our goal is to classify extragalactic time-varying sources into two broad classes, BL or SV, we assess the general shapes of the light curves by comparing their similarities to SN-like bursting behavior, or to AGN-like damped-random-walk type behavior. While fitting an exact model involves a large number of parameters which may be unknown, and may necessitate a large number of data points, the general shape of a BL light curve can be approximated to certain simpler analytical functional forms (Gaussian, Gamma distribution, and generic analytic SN model); and that of an SV light curve approximated by an OU process \citep{ornuhl1930,KimBailer-Jones2013} as described in Table \ref{Models}. In these models, we have ignored the effects of cosmological redshift corrections and dust extinction.  However this is acceptable since our goal here is to use the models only to distinguish between coherent single-burst type behavior from stochastic variability, while not assuming any underlying physical processes for the sources. Since the model parameters span several orders of magnitude, their priors are chosen to be uniform in the logarithm. 

\begin{figure}[htp]
\centering
\includegraphics[height=3in,width=3in]{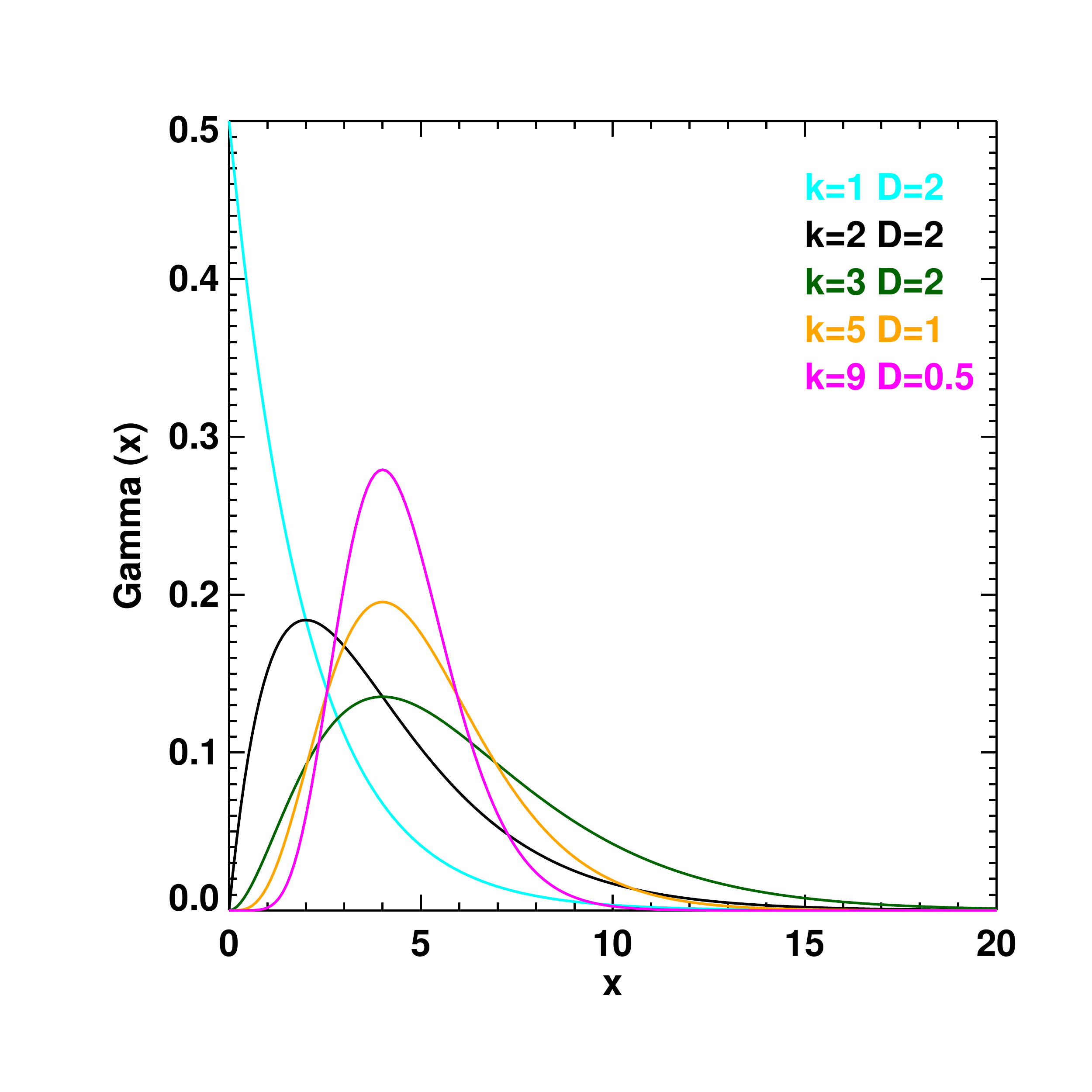}
\caption{A range of BL light curve shapes can be modeled using Gamma distributions by varying the shape and scale parameters $k,D$. This is particularly applicable to the asymmetric rise and fall time-series patterns of SN light curves.} \label{Gammafig} 
\end{figure}

The Gaussian is the simplest model that attempts to model the overall flux from a BL source as the sum of a constant background $\alpha$, and bursting behavior characterized by a Gaussian with amplitude $\beta$, center $\mu$, and width $\sigma$. This however, does not account for the asymmetry in SN light curves; for example Type Ia SNe are better approximated by a sharp rise $t_{\rm rise}\approx15$ days \citep{Gaitan2013,Ganeshalingam2011,Hayden2010} followed by a relatively slow decline in the flux in any band ($t_{\rm fall}\approx30$ days).  To resolve this, we employ a Gamma distribution which is robust in modeling such light curves (Fig.~\ref{Gammafig}), reflecting varying degrees of asymmetry depending on the shape $k$ and scale  $D$ parameters of the distribution. Another model is the Analytic-SN model, that uses distinct exponential rise and decline timescales, $t_{\rm rise}$ and $t_{\rm fall}$, and is particularly well suited to modeling non-Ia type SN light curves \citep{Kessler2010}, although it is generic in its application. Despite the non-specific nature of these models, we find that their simple statistical descriptions of the difference-flux light curves of BL sources are sufficient in distinguishing them from AGN with low contamination. The use of three distinct analytical BL models allows for a broader range of BL light-curve shapes, and is comparable to using independent statistical descriptions of the light curve through distinct parameterizations. Also, since the BL models are compared with the SV model, only their relative fitnesses in describing the data are important. Should the necessity arise of classifying the objects into particular sub-classes of the broader SV-BL distinction, or that of extracting particular details about the parameters of a source light curve, exact models for the sources \citep{Kessler2010,Kelly2010} must be included in the comparisons, which although it is beyond the scope of this paper, is a direct natural extension.  

The fluctuating behavior of optical light curves of AGN is well described by an OU process \citep{Kelly2009}, a first-order continuous-time auto-regressive process. The process can be described in terms of a driving noise field, parameterized by $c$, the square of the amplitude of the OU noise field, and a damping timescale $\tau$ \citep{Bailer-Jones2012}. Mathematically, the evolution of the state variable $Z(t)$ of the OU process is given by the differential equation 

\begin{eqnarray}
dZ(t) = c^{1/2} dW(t) - \frac{1}{\tau}(Z(t)-b) dt 
\end{eqnarray}

where $W$ is a Wiener process, and $b$ is the mean value of the process. For simulating the OU process itself, we use the prescriptions from \citet{Bailer-Jones2012}. The method uses posterior analysis to improve the estimate of the state variable continuously, by using the observed flux $y_{k-1}$ at time-step $t_{k-1}$ to compute the posterior distribution of the state variable $z_{k-1}$, which is subsequently used to update $z_k$. The OU process being a Gaussian-Markov process, $Z(t)$ is characterized by Gaussian probability distribution function $G(\mu(Z(t)),V(Z(t)))$ where $\mu,V$ are the mean and standard deviation of the state variable at time $t$. 

\begin{figure}[htp]
\centering
\includegraphics[scale=0.3]{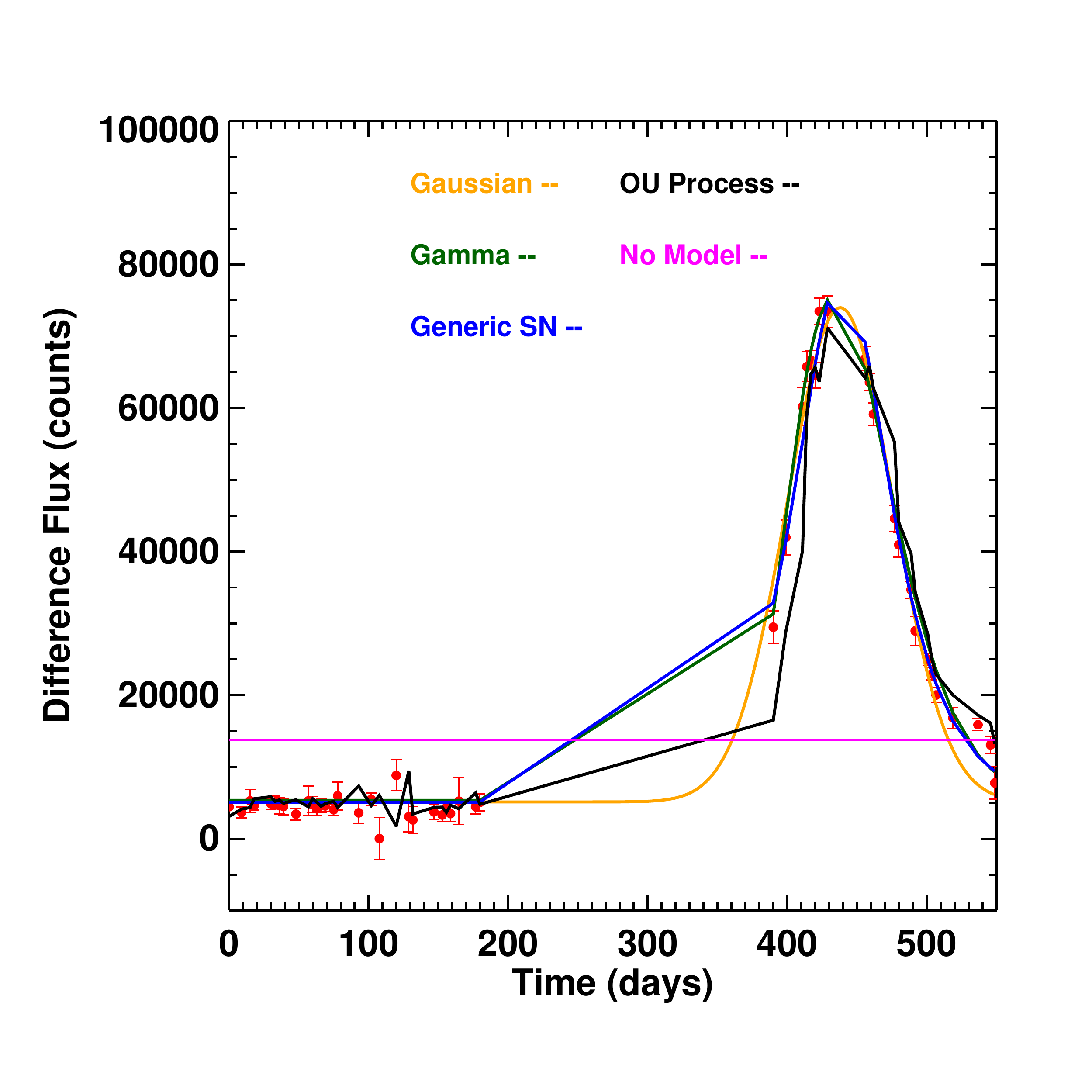}
\caption{An SN difference-flux light curve is reasonably well fit by all the models, but the BL models have higher LOOCV and lower AICc as compared to the OU process resulting in the light curve being classified as BL.} \label{SNEXAMPLE} 
\end{figure}

\begin{figure}[htp]
\centering
\includegraphics[scale=0.3]{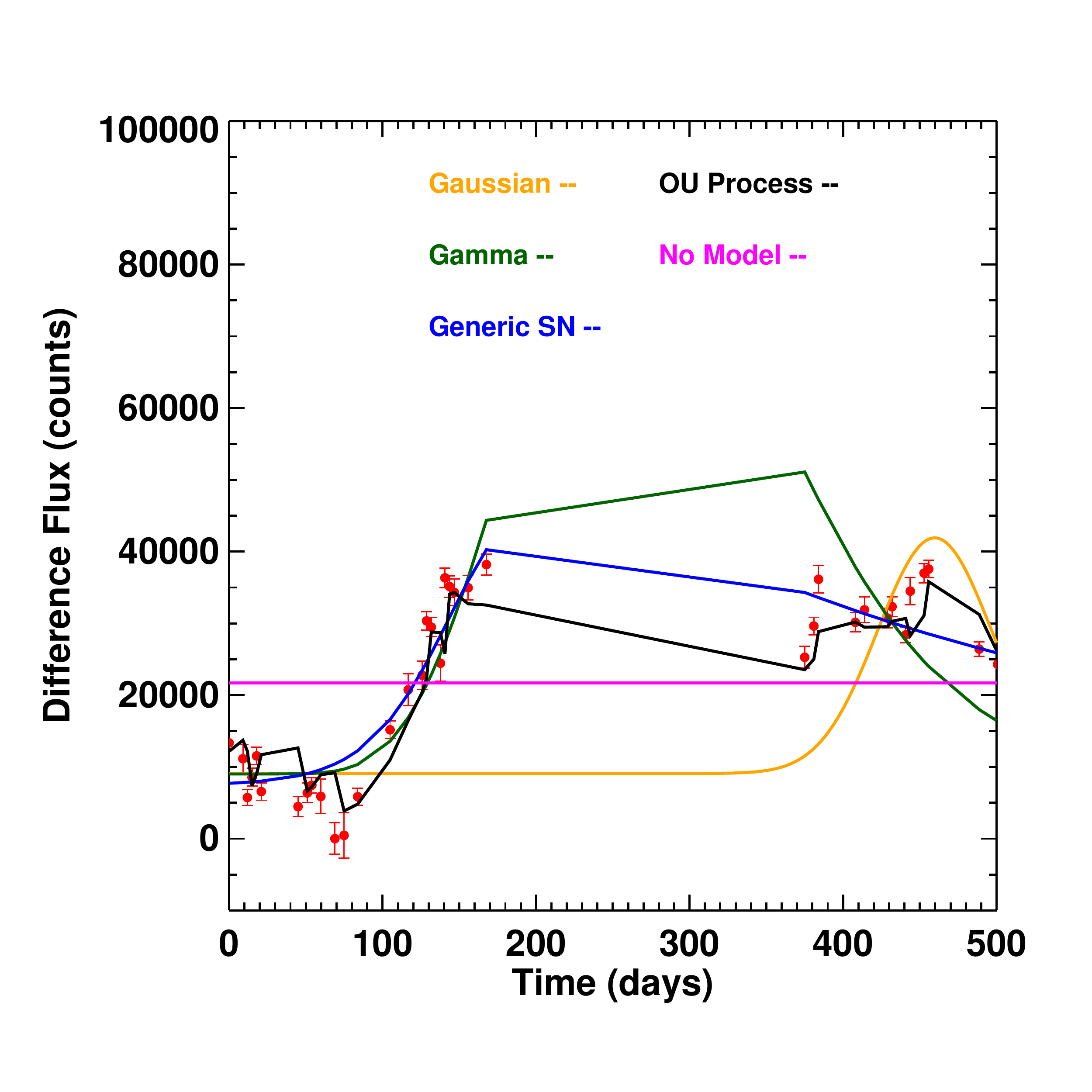}
\caption{Example of an AGN light curve that is well fit by the OU process and poorly by the BL models.} \label{AGNEXAMPLE} 
\end{figure}

\begin{figure}[htp]
\centering
\includegraphics[scale=0.3]{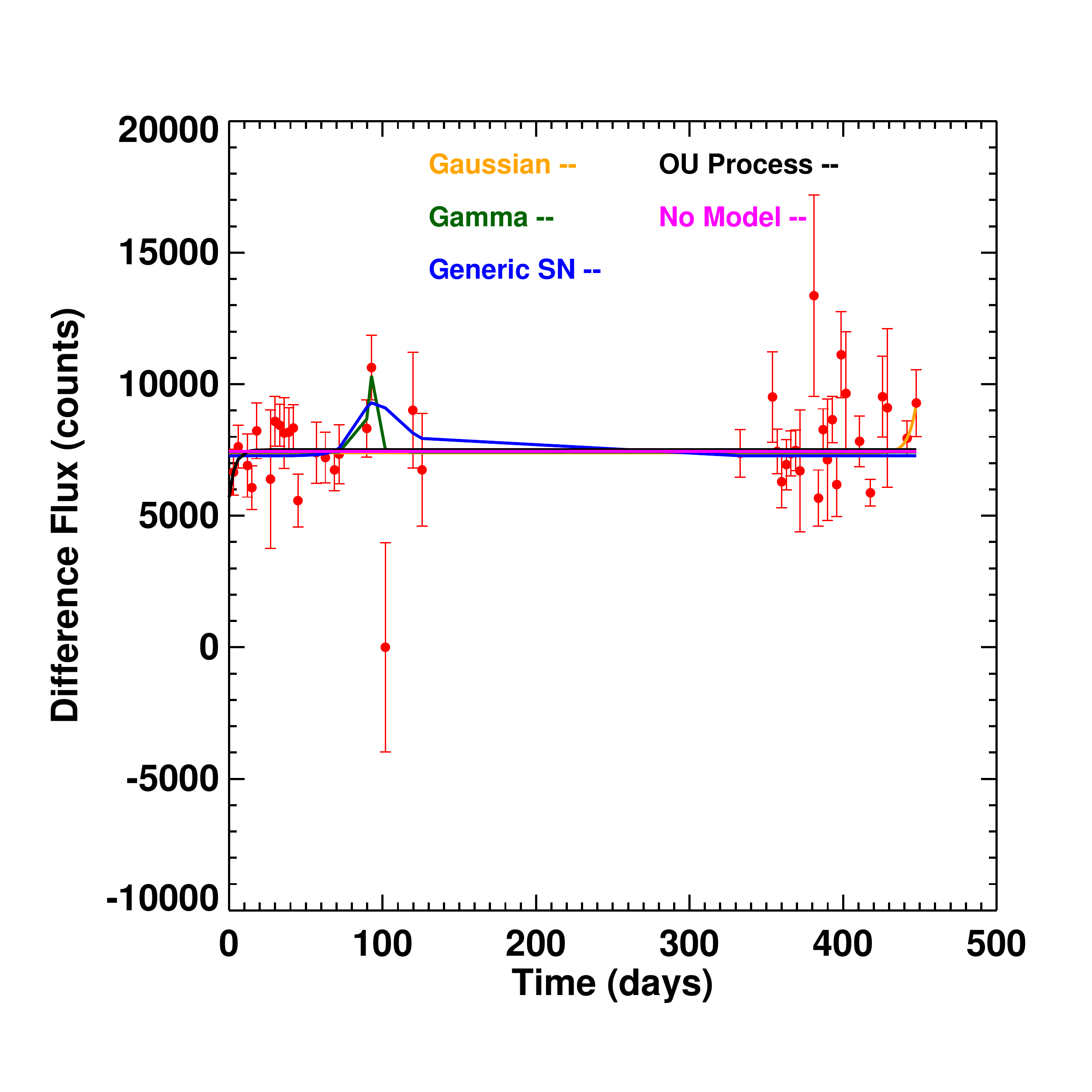}
\caption{Example of a difference-flux light curve that has a large number of image-differencing errors resulting in it being best fit by the No-Model.} \label{NMEXAMPLE} 
\end{figure}

We also determine whether the light curves are well fitted by a constant model that is representative of white noise. In the event that none of the light curve models is significantly better than white noise, the light curves are assumed to not pertain to any of the stochastic or bursting categories, and are classified as No-model (NM) sources. Figs.~\ref{SNEXAMPLE}, \ref{AGNEXAMPLE}, and \ref{NMEXAMPLE} show examples of SN, AGN, and NM classified sources and all the model fits. In each case, the best models are chosen based on robust statistical criteria, and the final source class decided using a clustering machine learning scheme described in the following sections. 

\begin{deluxetable*}{ccccc}
\tablecaption{Difference-flux models \label{Models}} 
\tablehead{\colhead{Model} & \colhead{Type} & \colhead{Equation} & \colhead{Parameters} & \colhead{Prior Distributions}}
\startdata
\hline
\rule{0pt}{4ex} Gaussian & BL & $Flux(t) = \alpha + \beta e^{-(t-\mu)^2/\sigma^2}$ & $\alpha$ & $Ulog(0,10^5) ({\rm flux~counts})$\\
& & & $\beta$ & $Ulog(10^3,10^8) ({\rm flux~counts})$\\
& & & $\mu$ & $Ulog(-10^2,10^4) ({\rm days})$\\
& & & $\sigma$ & $Ulog(1,10^4) ({\rm days})$\\
\hline
\rule{0pt}{4ex} Gamma distribution & BL & $Flux(t) = \alpha  + \beta \frac{(t-\mu)^{k-1} e^{-(t-\mu)/D}}{D^k \Gamma(k)}$ & $\alpha$ & $Ulog(0,10^5) ({\rm flux~counts})$\\
& & & $\beta$ & $Ulog(0,10^9) ({\rm flux~counts})$\\
& & & $\mu$ & $Ulog(-10^2,10^4) ({\rm days})$\\
& & & $D$ & $Ulog(1,10^2)$\\
& & & $k$ & $Ulog(1,10^2)$\\
\hline
\rule{0pt}{4ex} Analytic-SN model & BL & $Flux(t) = \alpha + \beta \frac{e^{-(t-t_o)/t_{fall}}}{1+e^{-(t-t_o)/t_{rise}}}$ & $\alpha$ & $Ulog(0,10^5) ({\rm flux~counts})$\\
& & & $\beta$ & $Ulog(0,10^9) ({\rm flux~counts})$\\
& & & $t_o$ & $Ulog(-10^2,10^4) ({\rm days})$\\
& & & $t_{fall}$ & $Ulog(1,10^3) ({\rm days})$\\
& & & $t_{rise}$ & $Ulog(1,10^3) ({\rm days})$\\
\hline
\rule{0pt}{4ex} OU process & SV & $dZ(t) = -\frac{1}{\tau} Z(t) dt + c^{1/2} \it{N}(t;0,dt)$ & $\tau$ & $Ulog(1,10^6) ({\rm days})$\\ 
& & where $Z$ here is flux count. & $c$ & $Ulog(0,10^{14}) ({{\rm flux~counts}}^2)$\\
& & & $b$ & $Ulog(0,10^8) ({\rm flux~counts})$\\
& & & $\mu(Z)$ & $Ulog(0,10^8) ({\rm flux~counts})$\\
& & & $V(Z)$ & $Ulog(0,10^{14}) ({{\rm flux~counts}}^2)$\\
\hline
\rule{0pt}{4ex} No-Model & White Noise & $Flux(t) = C$ & $C$ & $Ulog(0,10^8) ({\rm flux~counts})$ 
\enddata
\end{deluxetable*}

\section{Model Likelihood and Fitness Estimation}\label{FITNESS} 

For all time-series models, including for the OU process, we assume a Gaussian error model to compute the model likelihoods. Although for an OU process, the actual likelihood is computed differently from this \citep{Bailer-Jones2012}, using only the photometric errors to compute the likelihood for all models, is justified, since the intent is to determine the model-mean that best mimics the light curve shape. The probability $P(y_k | \sigma_k,\theta_n)$ of observing a difference flux $y_k$, assuming Gaussian errors, is given by    

\begin{eqnarray}
log P(y_k | \sigma_k,\theta_n) = log\left( \frac{1}{\sigma_k \sqrt{2\pi}}\right) - \frac{(f_k (\theta_n)-y_k)^2}{2\sigma_k^2}  \label{Lik}  
\end{eqnarray}

where $f_k$, $y_k$, and $\sigma_k$ are the model difference-flux, the observed difference-flux, and the standard deviation estimates of the $k$th data point. For the OU process we use $\mu(Z(t_k))$ or the mean light curve, in the place of $f_k$, to evaluate its likelihood. 

To assess the fitness of the models, we estimate their corrected Akaike information criteria (AICc) \citep{BurnhamAnderson2002} and leave-out-one cross-validation likelihoods (LOOCV) \citep{Bailer-Jones2012} over the difference-flux data for each source, filter-wise. The AIC (Eq.\ref{AICc}) is a quantification of the information lost when a model is used to represent a dataset. The AIC penalizes the maximum model log-likelihood $ln \mathcal{L}$ by a factor that depends on the number of model parameters $k$, thereby accounting for over-parameterization of the dataset. 

\begin{eqnarray}
AIC = 2k - 2ln \mathcal{L} \label{AICc}
\end{eqnarray}  

\begin{eqnarray}
AICc = AIC + \frac{2k (k+1) }{n-k-1} \label{AICceq}
\end{eqnarray}  

The AICc is a correction to the AIC, that corrects for the finite size of the dataset $n$ relative to the number of model parameters $k$. Note, that models that better represent the dataset have smaller AICc values. The LOOCV, another independent measure of model fitness, is a measure of how well each difference-flux value can be predicted using the remaining difference-flux data and hence, is a more robust statistical measure of model fitness as compared to the AICc. Re-stated, the cross-validation can also be said to measure the predictive ability of a model on a given dataset, and is a nearly unbiased estimator \citep{Kohavi} of the mean squared error for new observations. The LOOCV, more specifically, is the product of the piece-wise probability of obtaining individual difference-flux measurements using a time-series model, while sampling the parameters from the posterior constituted by the model over the remaining points in the time-series. In LOOCV estimation of a model over a dataset $y_k$ containing $K$ points, the likelihood $L_k$ of the $k$th data point is given by 

\begin{eqnarray}
L_k = P\left(y_k | y_{-k},\sigma,\eta\right) \approx \frac{1}{N} \sum_{n=1}^{n=N} P(y_k | \sigma_k,\theta_n,\eta) \label{LIK}
\end{eqnarray}

where $\eta$ is the time-series model, $\sigma_k$ is the error estimate at each point, and $\theta_n$ are the model parameters drawn in the $n$th iteration from the Markov chain Monte-Carlo sampling of the posterior probability distribution of the model over the other $K-1$ data points denoted by $y_{-k}$. The LOOCV of the model can then be obtained by multiplying the partition probabilities

\begin{eqnarray}
LOOCV = \prod_{k=1}^{k=K} L_k
\end{eqnarray}  

Since the AICc and the LOOCV are measured independently of each other, they can be used simultaneously to assess model likelihood, thereby reinforcing model fitness assessment. The LOOCV for each model is estimated using a Markov chain Monte-Carlo (MCMC) using a standard Metropolis-Hastings algorithm to sample the posterior distributions \citep{Hastings1970}. The model parameters are sampled from known distributions and the posterior probability $L_i p_i$ is evaluated, where $p_i$ is the prior probability and $L_i$ is the model likelihood in the $i$th iteration. Parameters for the $i+1$th iteration are accepted with probability $(L_{i+1}p_{i+1})/({L_i}p_{i})$, failing which the parameters from the $i$th iteration are retained. We use a log-normal sampling distribution with a diagonal covariance matrix, with $\sigma^2_{ii}=10^{-4}$ uniformly across all parameters, and all models. We find that this choice of a constant variance of the sampling distribution leads to stable cross-validation likelihood values. In accordance with log-normal sampling requirements, the  parameter distributions defined in Table \ref{Models} are transformed between $-\infty,\infty$ using a sigmoidal transform. 

The prior distributions for the BL and SV model parameters can be assumed to be uniform in the logarithm, as we have in our simulations, or can be obtained by sampling the parameters at the posterior maxima for the BL and SV models, for known SNe (BL) and AGN (SV) training sets. The latter is advantageous if the entire set of sources is well represented by the training set, in that the number of iterations to convergence would be significantly reduced. However, we did not make this assumption in order to allow for the classification of BL and SV light curve types that may not occur in the verification set, and only took care to ensure that the limits on the parameter ranges subsumed the parameter values that could occur in the dataset. 

Since the initial guesses for the model parameters in the MCMC may be far from the actual solution, a burn-in of $1000$ iterations is employed for all model assessments. We determined that a large number of burn-in iterations is important to ensure sampling near the peaks of the posterior distribution, and is particularly important while using prior parameter distributions that are uniform in the logarithm, as we have done here. We determined that $10000$ post burn-in iterations were sufficient for good model-fit convergence, after replicating the results with $2000$ burn-in iterations, and $20000$ post burn-in iterations. To ensure that the Markov chains were well mixed, we further computed the Gelman-Rubin statistic \citep{gelman1992} over 10 parallel chains, for each parameter in each model, and ensured that their values were less than $1.1$. The calculation of the LOOCV is tedious and computationally expensive, and required us to parallelize our codes over a $300$ core multi-node cluster. In addition our codes were written ground-up in C++ and optimized for quick run-time. The classification of $\approx7000$ sources with $\approx40$ difference-flux points in each of the four bands, required $\approx4$ hours.  

\section{Classification Method} \label{Classification}

Once the fitnesses of the models are estimated filter-wise on the difference-fluxes using the AICc and the LOOCV, we obtain two parameters per model for $5$ time-series models in each of the four filters $g_{P1},r_{P1},i_{P1},$ and $z_{P1}$.  First, we remove the NM best fit sources by comparing their model statistics with those of the BL and SV models. To do this we construct a relative sign vector $RV_{i,f}$ for each source, in each filter, using the AICc and the logarithms of the LOOCVs (which we designate as LLOOCV):    

\begin{eqnarray}
& &RV_{i,f}=\{\nonumber \\ 
& &sgn(LLOOCV_{Gaussian,i,f}-LLOOCV_{NM,i,f}),\nonumber \\
& &sgn(LLOOCV_{Gamma,i,f}-LLOOCV_{NM,i,f}),\nonumber \\ 
& &sgn(LLOOCV_{Analytic,i,f}-LLOOCV_{NM,i,f}),\nonumber \\
& &sgn(AICc_{Gaussian,i,f}-AICc_{NM,i,f}),\nonumber \\
& &sgn(AICc_{Gamma,i,f}-AICc_{NM,i,f}),\nonumber \\
& &sgn(AICc_{Analytic,i,f}-AICc_{NM,i,f}\nonumber\\
& &sgn(LLOOCV_{OU,i,f}-LLOOCV_{NM,i,f}),\nonumber \\
& &sgn(AICc_{OU,i,f}-AICc_{NM,i,f})\nonumber \\
& &\} \label{RV}
\end{eqnarray} 

where $i$ is the object id, $f$ is the filter, and $sgn$ denotes the sign function, defined to be $+1$ for positive values and $-1$ for negative values. Ideally, for a BL source, $RV_{BL} = \{+1, +1, +1, -1, -1, -1,\pm1,\mp1\}$ since the BL models will have a larger LLOOCV, and a smaller AICc when compared to the same for the NM, while for an SV source the relative sign vector should be $RV_{SV} = \{\pm1, \pm1, \pm1, \mp1,\mp1, \mp1,+1,-1\}$, i.e., the OU process better describes the light curve as compared to any of the BL models or the SV models. For sources where the NM is the best model, $RV_{NM} = \{-1,-1,-1,+1,+1,+1,-1,+1\}$. 

We then compute $RV_{i,f}$ for all sources, which are aggregated in filter-wise and fed into a K-means clustering supervised-machine-learning algorithm, using the number of centers $K=3$ in a swap method that is repeated over $100$ iterations \citep{Kanungo}. The clustering algorithm partitions the sources in the 8-dimensional $RV_{i,f}$ space,  into Voronoi cells to determine the centers of the distributions for BL, SV, NM, by minimizing the sum of squares of the distances of points $x_l$ within cluster $S_m$ from the means of the clusters $\mu_m$ that correspond to the different classes of sources:  

\begin{eqnarray}
\sum_{l=1}^{k} \sum_{x_l \epsilon S_m} ||x_l-\mu_m||^2
\end{eqnarray}         

Each source is then assigned a class $C_{i,f}$ as BL, SV, or NM depending on the center it is clustered around. The squared-distance of each source point $i$ in filter $f$, $D_{i,f}=|x_i-\mu_{C,f}|^2$ from the clustering center $\mu_{C,f}$ is a measure of how reliably it is classified as the particular type $C$, with a distance of $D_{i,f}=0$ being the best, and larger distances indicating less reliable classifications. $C_{i,f}$ and $D_{i,f}$ are computed for each source, in each of the $g_{P1},r_{P1},i_{P1},$ and $z_{P1}$ bands independently. We choose to classify the sources filter-wise, and not using the statistical measures from all the filters at once, for the following reasons.

\begin{enumerate}

\item The behavior of each type of source, across the filters, cannot be assumed to be uniform and hence, the clustering centers may differ significantly, 

\item Clustering in some filters may be more noisy than others, resulting in most sources being classified as no-model sources, thereby making these bands less favorable for classification purposes. For these filters, the no-model center would be repeated in place of an SV or a BL center. 

\item Some filters may be less noisy and show clustering only around two centers corresponding to BL and SV. Combining these filters with the noisier ones results in both, more uncertainty in clustering classification (larger $D_i$) and a larger number of misclassifications. This is because, the uncertainty in the clustering classification caused by one or more filters confounds the otherwise clear classifications from the others. As a result, the clustering centers are poorly determined in the joint parameter space of statistical parameters from all the filters. By performing the clustering in each filter separately, the classifications can be reinforced if they show agreement across filters, and reflect the uncertainty otherwise, via smaller $|C_i|$ and larger $D_i$ values. 

\end{enumerate}

Note, that it is favorable to assume more clustering centers in any filter than there are. For example, we could assume that a certain filter has $3$ clustering centers corresponding to BL, SV, or NM while it may so happen that one of the BL or SV centers is repeated, or two of the centers are relatively proximal, implying that the clustering really occurs only around two centers. Post clustering, we filter out sources which have been clustered around NM centers in at least $3$ bands they are detected in, and label them NM sources. The remaining sources then, are classified in at least $2$ bands they are detected in as either SV or BL. To detect their type more precisely than just using their comparisons to the no-model, we construct another relative sign vector $BLSV_{i,f}$ comparing the fitness statistics of the BL models directly to those of the OU process for each source, band-wise:  

\begin{eqnarray}
& &BLSV_{i,f}=\{\nonumber \\ 
& &sgn(LLOOCV_{Gaussian,i,f}-LLOOCV_{OU,i,f}),\nonumber \\
& &sgn(LLOOCV_{Gamma,i,f}-LLOOCV_{OU,i,f}),\nonumber \\ 
& &sgn(LLOOCV_{Analytic,i,f}-LLOOCV_{OU,i,f}),\nonumber \\
& &sgn(AICc_{Gaussian,i,f}-AICc_{OU,i,f}),\nonumber \\
& &sgn(AICc_{Gamma,i,f}-AICc_{OU,i,f}),\nonumber \\
& &sgn(AICc_{Analytic,i,f}-AICc_{OU,i,f}\nonumber\\
& &\} \label{BLSV}
\end{eqnarray} 

We aggregate $BLSV_{i,f}$ band-wise and perform a two-center K-means clustering $(K=2)$ to segregate the BL and SV sources.  We find that this type of hierarchical supervised clustering, i.e. filtering out the NM sources in the first stage and classifying the remaining sources as BL or SV in the second stage, is more efficient as compared to using all the model statistical comparisons concurrently in a single clustering step. This is because, model comparisons which are not relevant to a particular classification type, contribute significantly to the noise in the clustering process. We also attempted a clustering on the differences between the model LLOOCVs and AICcs, instead of on the signs of their differences, in a single clustering step, as well as in a hierarchical supervised method as discussed in this paper.  However, we found that in both cases, the number of misclassifications is larger due to the associated variance in the values of the differences in LLOOCV and AICc, which is mitigated by reducing them to binary statistics using the sign of their differences alone. 

We combine the clustering classifications from the second clustering stage in each filter, by defining two measures; a quality factor $C_i$ which is the average of classifications across the filters:  

\begin{eqnarray}
C_i = \frac{\sum_f C_{i,f}}{N_{filters}} \label{class} 
\end{eqnarray}  

and, the average clustering square distance $D_i$ across the filters:  

\begin{eqnarray}
D_i = \frac{\sum_f D_{i,f}}{N_{filters}} \label{distclass}
\end{eqnarray}

\begin{figure*}
\centering
\subfloat[]{\includegraphics[scale=0.3]{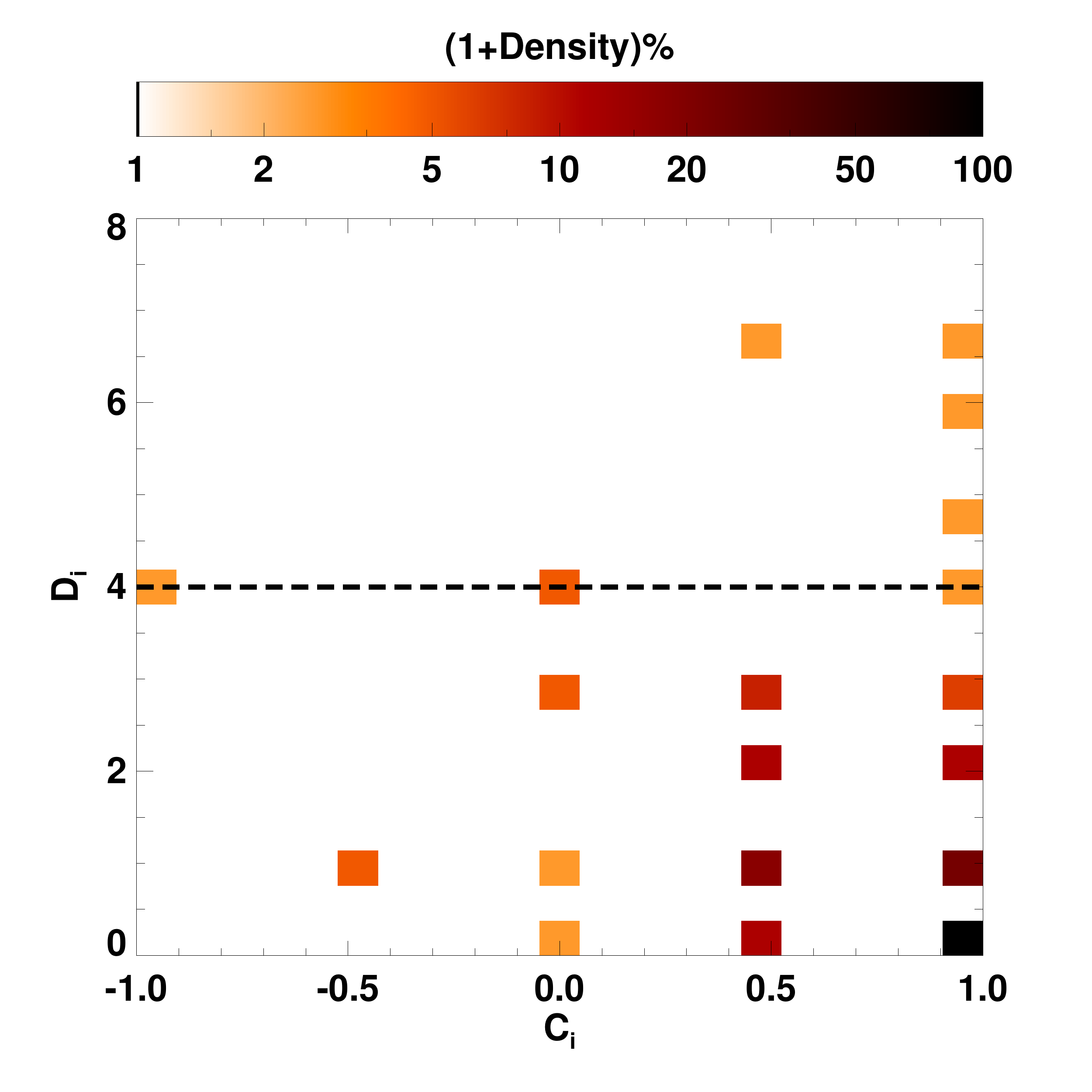}}
\subfloat[]{\includegraphics[scale=0.3]{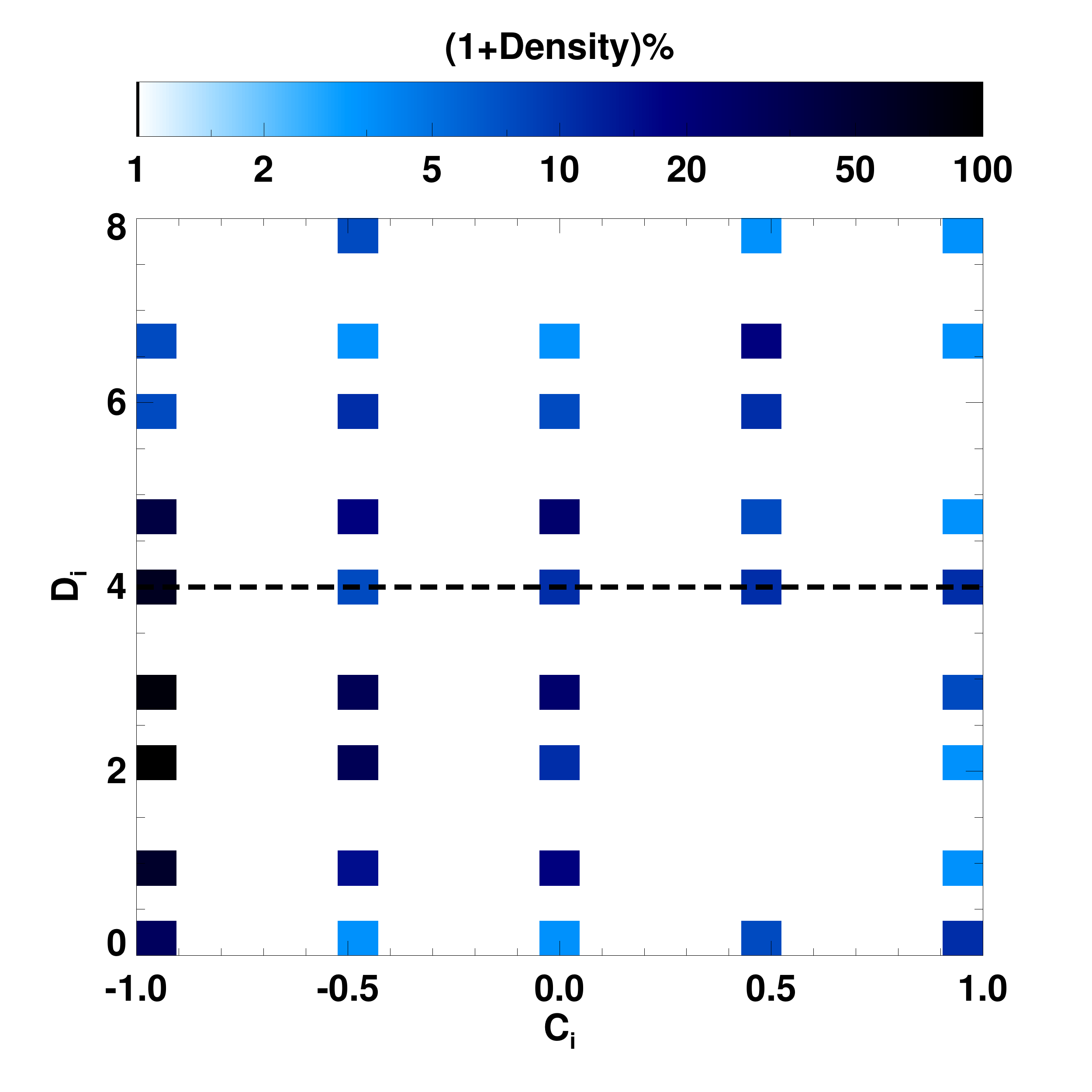}}
\caption{Densities of verification set SNe (left) and AGN (right) on the $C_i$ vs $D_i$ plane. Since AGN classifications for $D_i>4$ occupy both the BL ($C_i>0$) and SV($C_i<0$) regions, we only rely on classifications with $D_i<=4$. As a result, the SNe are classified with $93.89\%$ completeness and $90.97\%$ purity while the AGN are classified with $57.92\%$ completeness and $95.00\%$ purity. }
\label{AGNSNdist}
\end{figure*} 

BL sources have $C_i$ closer to $1$ while stochastically variable sources have $C_i$ close to $-1$. $D_i$ is a measure of the overall reliability of the classification that decreases with increasing $D_i$.  Therefore, sources which are purely BL will have $C_i=1, D_i=0$, while purely stochastic variables will have $C_i=-1,D_i=0$. Intermediate values of $C_i$ indicate disagreements between some of the band-wise classifications, while larger values of $D_i$ indicate a disagreement between the models in a given band. 

\subsection{Tests On a Verification Set}

To test our classification method we constructed a reliable verification set with a diverse range of SNe (BL) and AGN (SV) in order to capture, as much as possible, the full range of their time-variability properties. For AGN, we created a verification set from two sources: 1) $58$ UV-variability selected AGN with associations with PS1 alerts within $1 \arcsec$ from the GALEX Time Domain Survey (TDS) \citep{Gezari2013} with no available spectroscopy, and 2) $125$ spectroscopically confirmed AGN PS1 alerts associated with galaxy hosts from SDSS\citep{Shen2011} and from a multipurpose Harvard/CfA program with the MMT to observe PS1 transients (PI Berger). The GALEX AGN were selected from UV variability at the $5\sigma$ level in at least one epoch, and then classified using a combination of optical host colors and morphology, UV light curve characteristics, and matches to archival X-ray, and spectroscopic catalogs. 
The SN verification set consists of $131$ spectroscopically confirmed Type-Ia, Type-Ib$/$c, Type-II, Type-IIn, and Type-IIP SNe from a combination of PS1 spectroscopic follow-up programs using Gemini, Magellan, and MMT described in \citet{Rest2013} and Berger et al. (in prep.).  In order to test the performance and efficiency of our algorithm in the classification of AGN and SNe, we have constructed a diverse and robust verification set that should be representative of these populations in our sample. 

\begin{figure*}[htp]
\centering
\includegraphics[width=7in,height=6in]{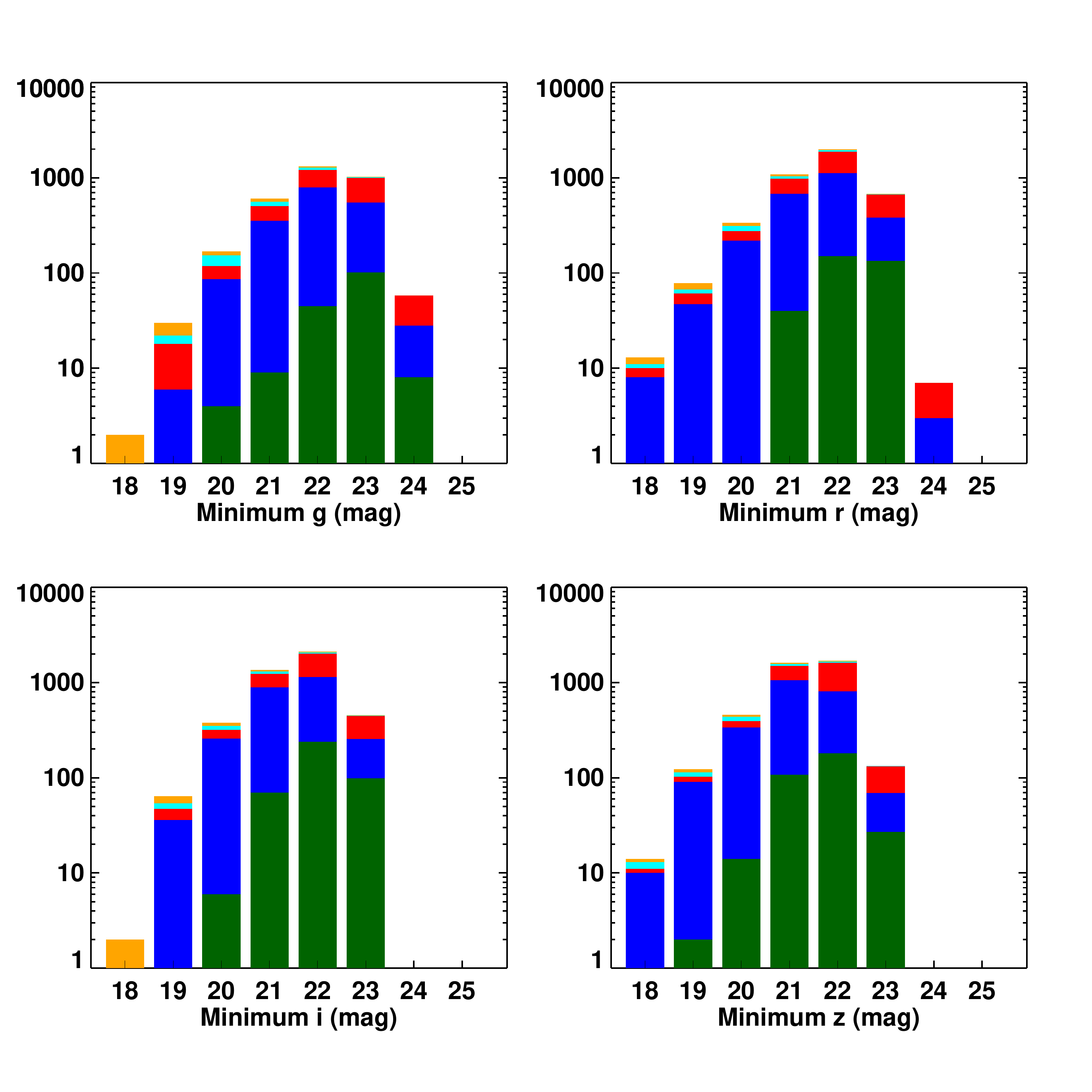}
\caption{Distribution of SV (blue), verification set AGN (cyan), BL (red), verification set SN (orange), and NM (dark green) as a function of minimum magnitudes in the $g_{P1},r_{P1},i_{P1}$, and $z_{P1}$ bands. } \label{NOMODELMINMAG} 
\end{figure*}

\begin{figure}[htp]
\centering
\includegraphics[scale=0.3]{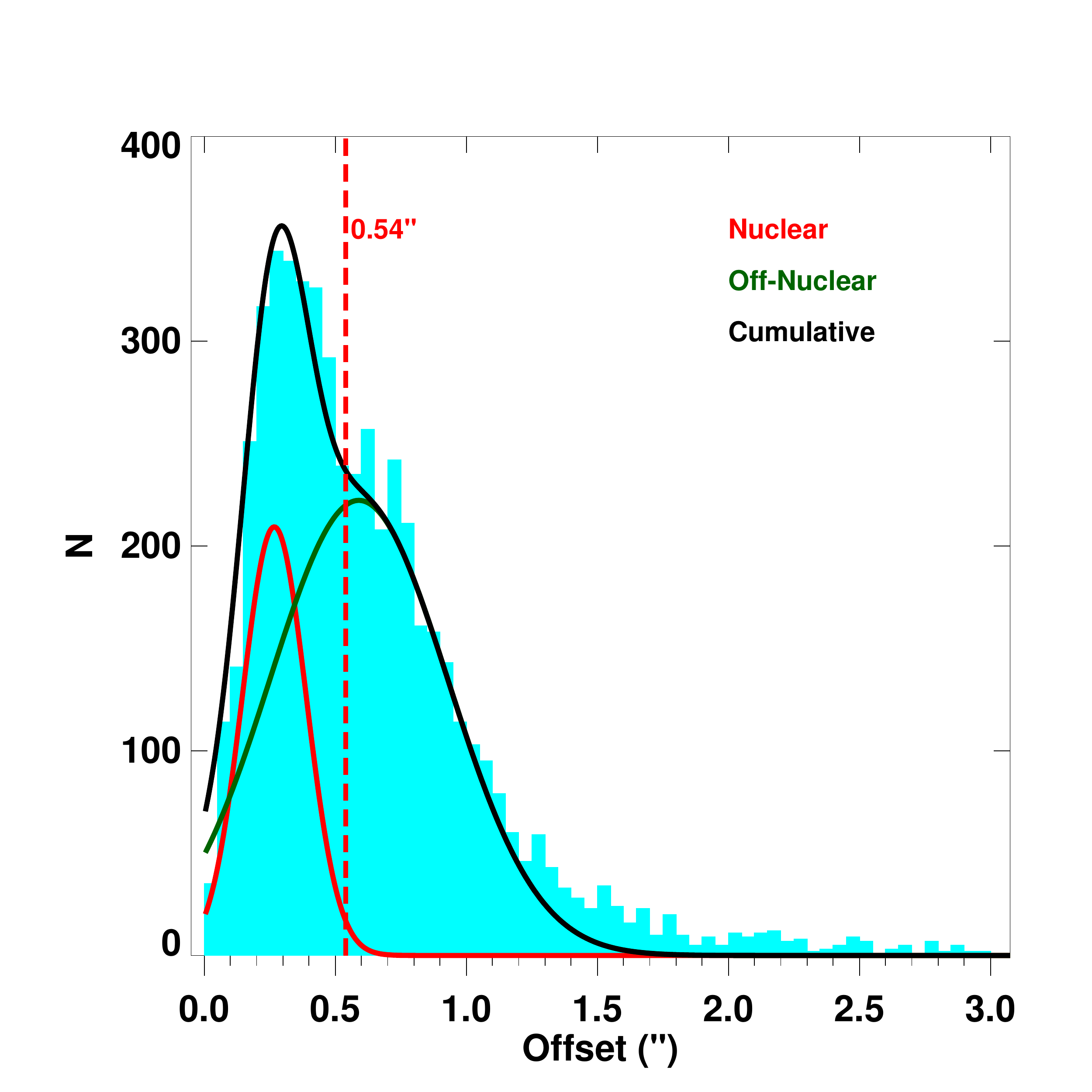}
\caption{Bimodal distribution of PS1 extragalactic alert host offsets, separated into nuclear, and off-nuclear distributions, with $\mu_{\rm nuc},\sigma_{\rm nuc}=0.26,0.14$ and $\mu_{\rm off-nuc},\sigma_{\rm off-nuc}=0.48,0.37$. Sources offset from their host galaxies by more than $\mu_{\rm nuc}+2\sigma_{\rm nuc}=0.54"$ are predominantly SNe. } \label{offset} 
\end{figure}

\begin{figure}[htp]
\centering
\includegraphics[scale=0.3]{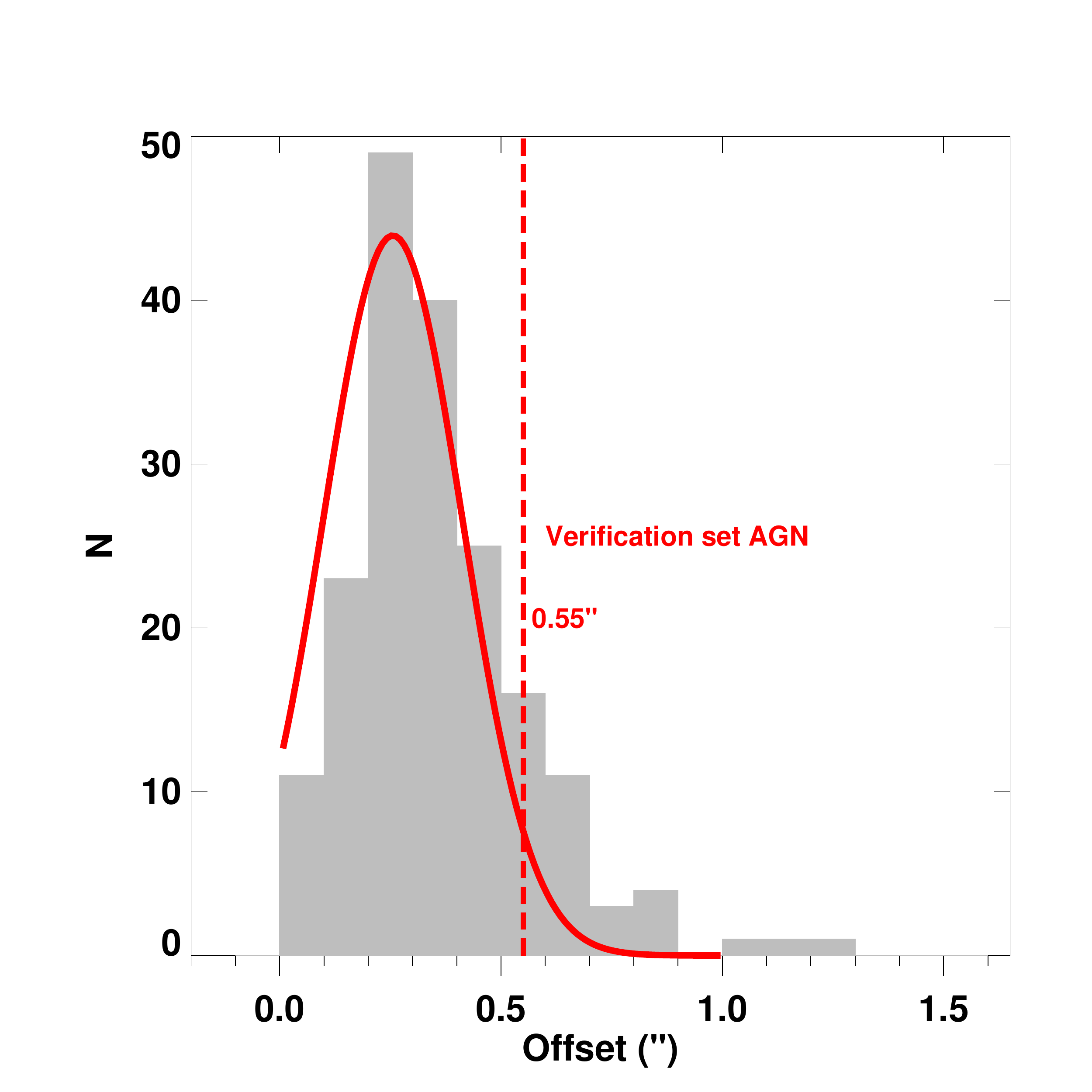}
\caption{Offset distribution of verification set AGN which is well approximated by a Gaussian with $\mu_{AGN},\sigma_{AGN}=0.25,0.15$. This is approximately the same as the distribution for nuclear offsets obtained in Fig.\ref{offset} from the bi-modal assumption for the entire extragalactic alerts population.} \label{AGNGALEXoffset} 
\end{figure}

Fig.~\ref{AGNSNdist} shows the contours of $C_i$ vs $D_i$ for the spectroscopic AGN and SNe. The SNe cluster around the region $C_i\geq0.5$ and $D_i<4$, while AGN  predominantly occupy the regions defined by $C_i\leq0$ and $D_i\leq8$. In general, the degree to which an object is BL as opposed to SV increases with $C_i$. Some AGN light curves may show bursting-type behavior resulting in their being classified in more than $1$ filter as BL, consequently having $C_i>-1$. Also, AGN clustering are less reliably classified as evidenced by systematically larger $D_i$ as compared to the SNe. 

This is possibly because, the damped random walk model may be over-simplifying the description of the underlying AGN variability. A more complex model such as in \citet{Kelly2010} or \citet{KellyBecker2014}, may be required to capture the true variability. Another source of confusion is that some of the AGN may resemble BL light curves if they fluctuate above the detection threshold for only a brief period during the observing baseline.  Consequently, $47.88\%$ of the verification set AGN classified with $D_i>4$, indicating unreliable classifications. In order to maximize the purity of our classifications, with a sacrifice to completeness, we use  $D_i=4$ as the bound for the classifications, below which $57.92\%$ AGN and $93.89\%$ SNe, are recovered with $95.00\%$  and $90.97\%$ purities respectively. It is possible to include other photometric properties like color or host-galaxy properties to improve the completeness of the AGN classifications, however, since the focus of our present work is to only use time-variability as a tool for classification, we reserve this for future work.

\begin{figure}[htp]
\centering
\includegraphics[scale=0.3]{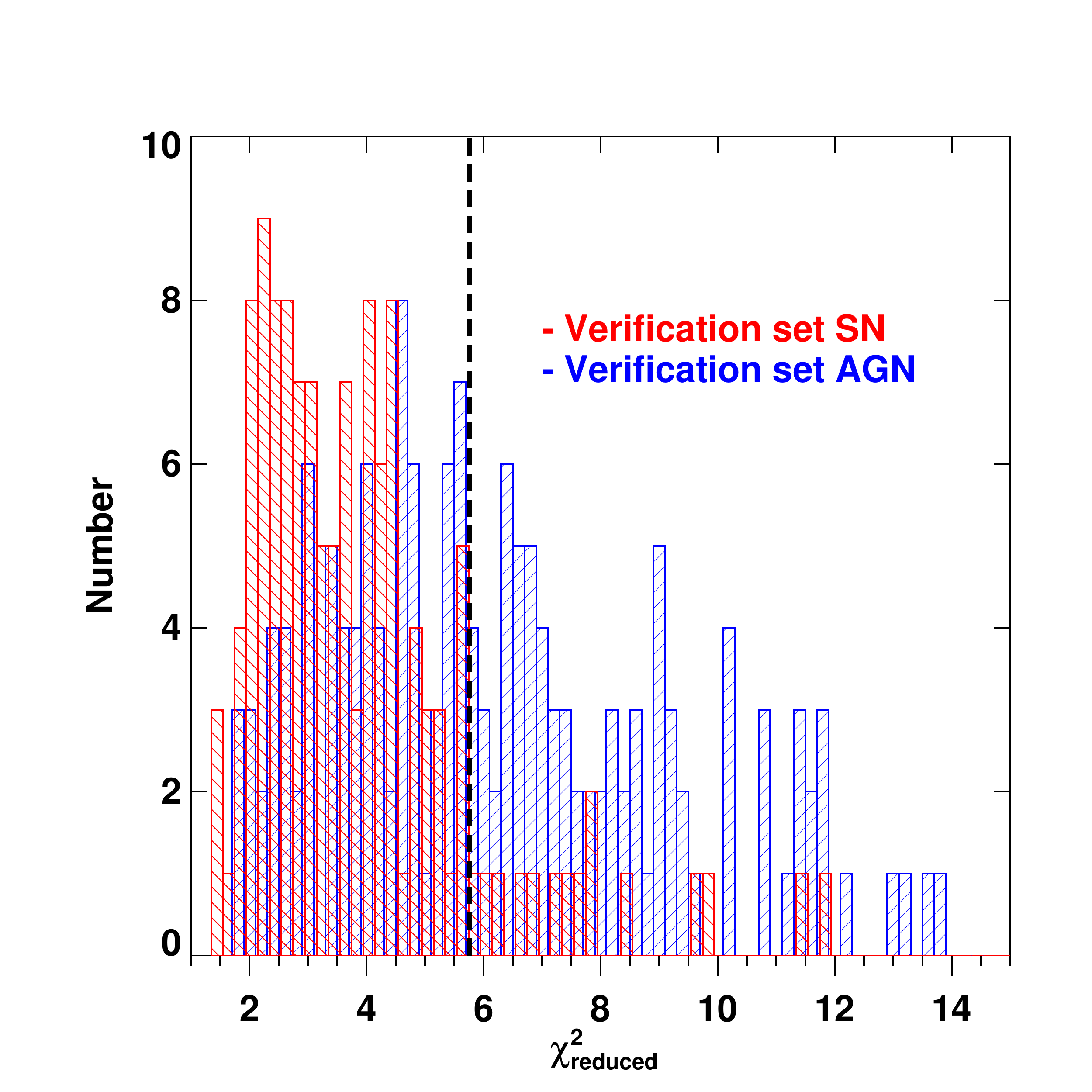}
\caption{Minimum seasonal $\chi^2_{\nu}$ of the spectroscopic verification set of AGN and SNe in the $g$ band. A cut-off of $\chi^2_{\nu}>5.75$ can be used to demarcate AGN from SNe, albeit with a high contamination rate for SNe.} \label{SPECREDCHI2} 
\end{figure}

In multi-epoch surveys such as Pan-STARRS1, it may be possible to differentiate between AGN and SNe simply by comparing variability between observing seasons.  For example, an SN will have only one season in which the reduced $\chi^2$ ($\chi^2_{\nu}$) is larger than 1, while an AGN light curve can show variability in both seasons with $\chi^2_{\nu} >>1$.  In Fig.\ref{SPECREDCHI2} we test this simplified method by plotting the minimum of the seasonal $\chi^2_{\nu}$s for the verification set AGN and SNe. If $\chi^2_{\nu}=5.75$ is used to separate AGN from SNe, then $55.73\%$ of the AGN can be recovered with $86.45\%$ purity. We find that, using this value of $\chi^2_{\nu}$ to separate AGN and SNe, results in maximal purity for the AGN sample. While this method yields a completeness for AGN that is similar to that of our light-curve classification algorithm, for SNe, the performance is much worse, with 87.69\% of SNe recovered at 47.22\% purity.  This high contamination rate for SNe is due to the extensive overlap between the AGN and SN in the region $\chi^2_{\nu}<5.75$.  Hence, we conclude that for maximum purity, a more sophisticated method such as the one adopted in this paper is necessary. 

\subsection{Final Classifications and Properties of Extragalactic Sources}

We begin classifying our $4361$ extragalactic transient alerts by first selecting out sources which are clustered around the NM center in at least $3$ of the $4$ bands. We find $570$ such sources (NM sources hereafter).    Visual inspection of the NM source light curves reveals that the majority are the result of noisy image-differencing light curves, most often due to large excursions in flux from image differencing artifacts, and not statistical errors due to faint fluxes, in most of the bands.  This can be seen in Fig.~\ref{NOMODELMINMAG}, which shows the minimum source magnitude in the $g_{P1},r_{P1},i_{P1}$, and $z_{P1}$ bands for all sources, including that for NM sources (dark green), which barring the brightest end of the magnitude distribution, follows the overall magnitude distribution of extragalactic sources (black), indicating no strong biases toward fainter magnitudes.


\begin{deluxetable*}{ccc}
\tablecaption{Source variability and offset classifications \label{Classifications}}
\tablehead{\colhead{Type} & \colhead{Nuclear (offsets $< 0.55\arcsec$)} & \colhead{Off-Nuclear (offsets $> 0.55\arcsec$)}}
\startdata
\hline
\rule{0pt}{2ex} Burst-Like & $689$  & $812$ (SNe)  \\
Stochastic Variable & $1233$ (AGN) & $1027$ \\
No-Model & $449$ & $121$ 
\enddata
\end{deluxetable*}

\begin{figure*}
\centering
\subfloat[]{\includegraphics[scale=0.3]{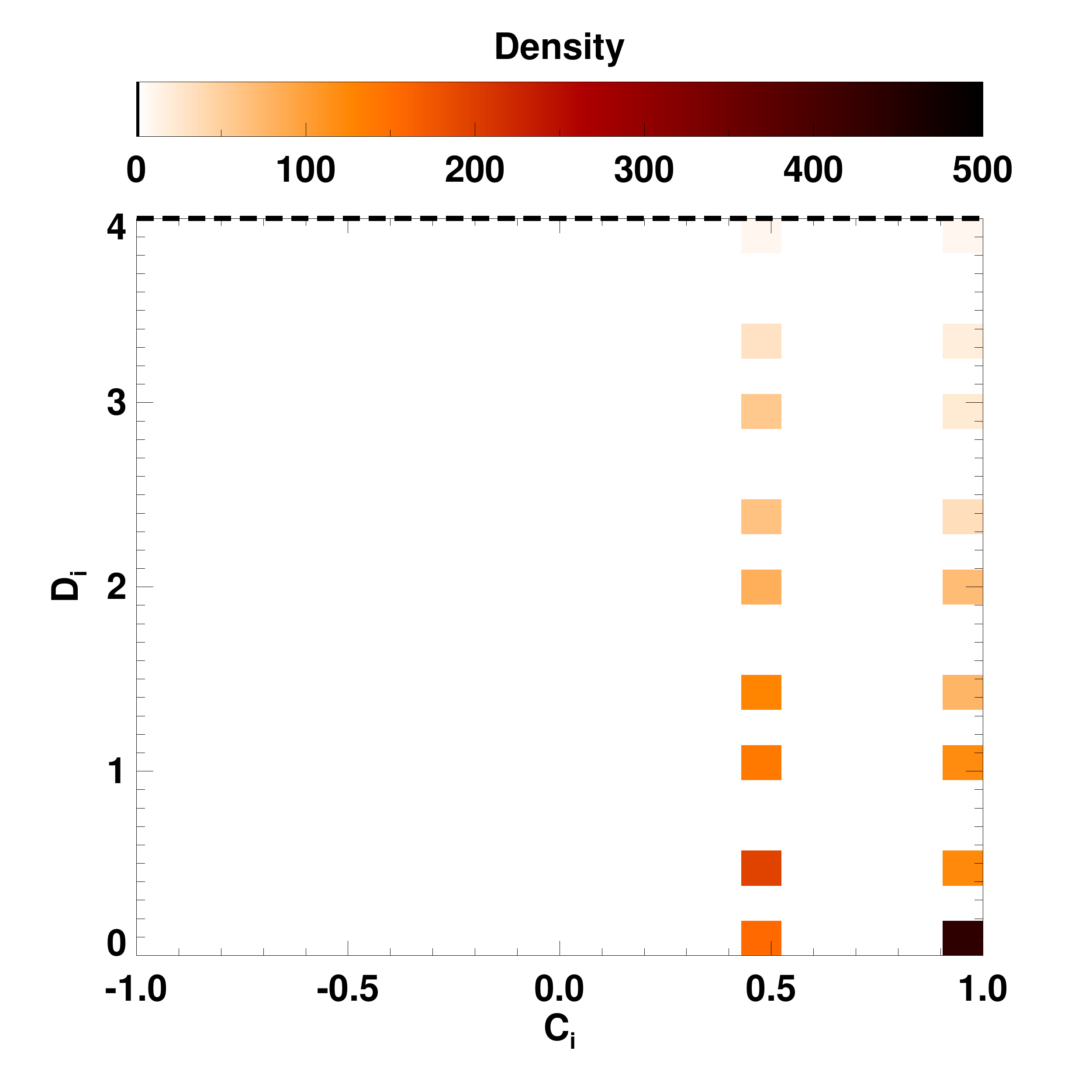}}
\subfloat[]{\includegraphics[scale=0.3]{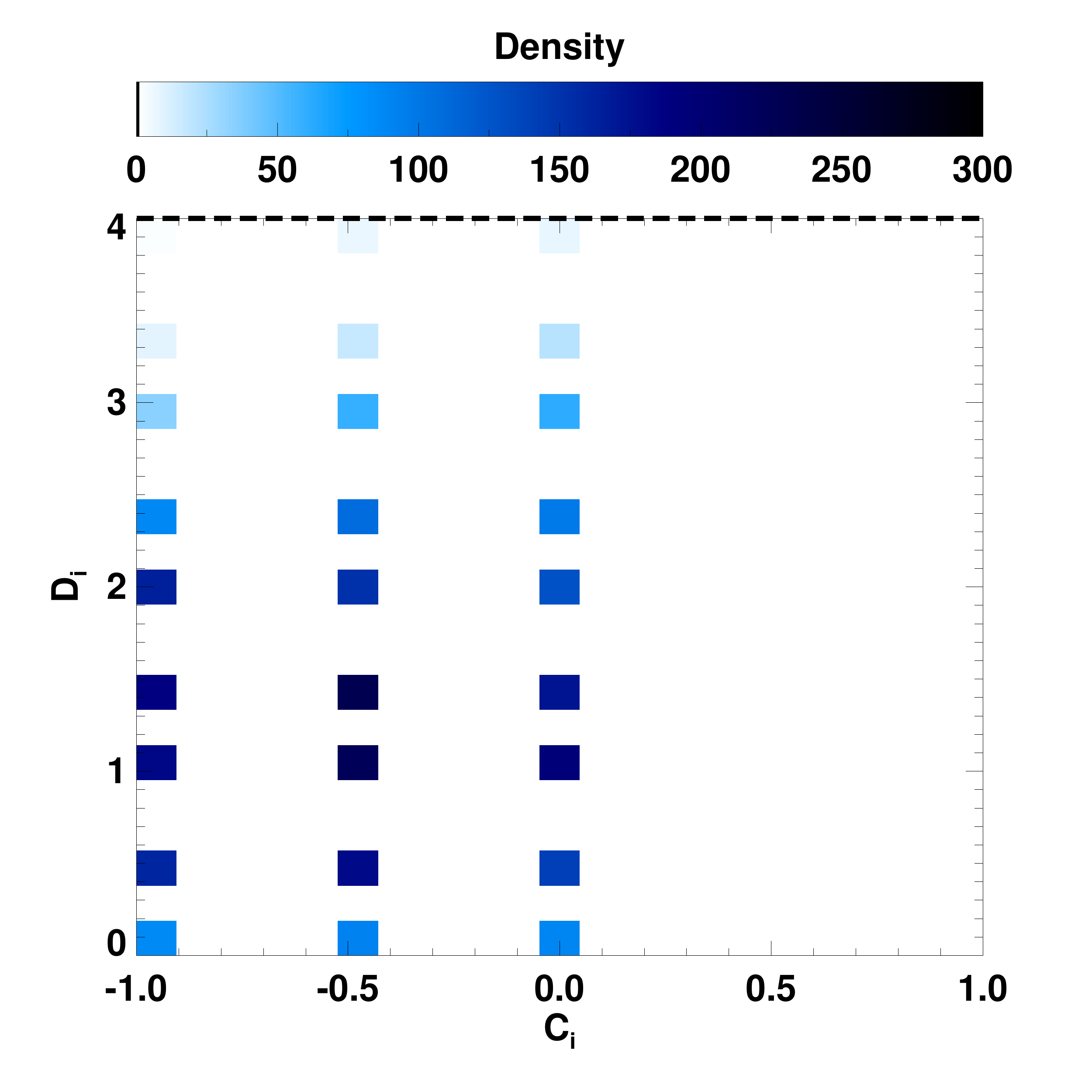}}
\caption{Density maps for BL (left) and SV(right) as a function of $C_i$ and $D_i$.}
\label{AllObj}
\end{figure*}

\begin{figure*}[htp]
\centering
\includegraphics[width=7in,height=5in]{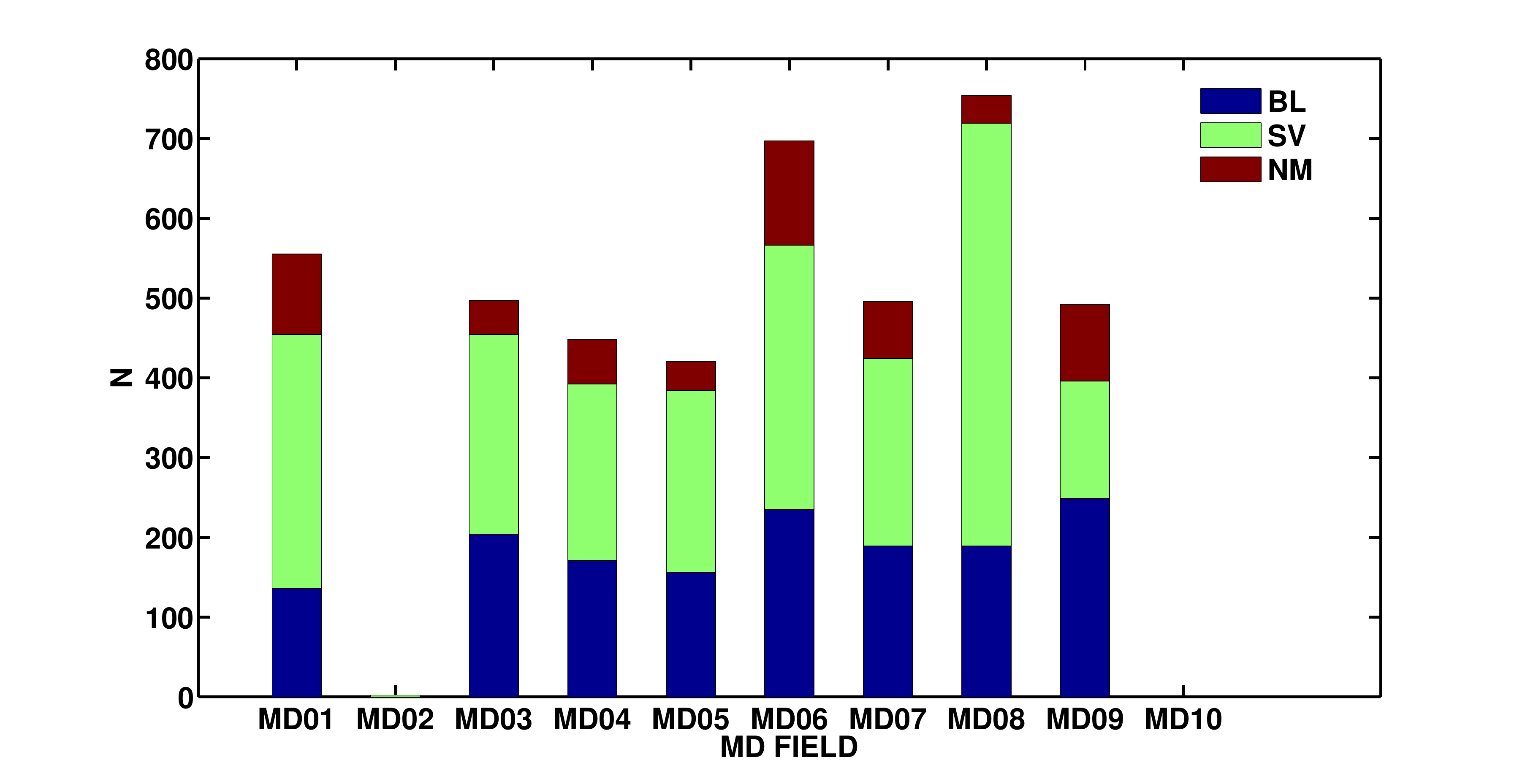}
\caption{Distribution of SV,BL, and NM sources across the $10$ MD fields.} \label{ALLFIELD} 
\end{figure*}

Fig.~\ref{AllObj} shows $C_i$ vs $D_i$ contours for the $3791$ extragalactic sources classified SV and BL. We determine that there are $2262$ SV sources and $1529$ BL sources in the dataset.  Fig.~\ref{ALLFIELD} shows the distributions of SV, BL, and NM by MD field, with SV being the most common class of extragalactic alert in all fields.  

\begin{figure}[htp]
\centering
\includegraphics[scale=0.3]{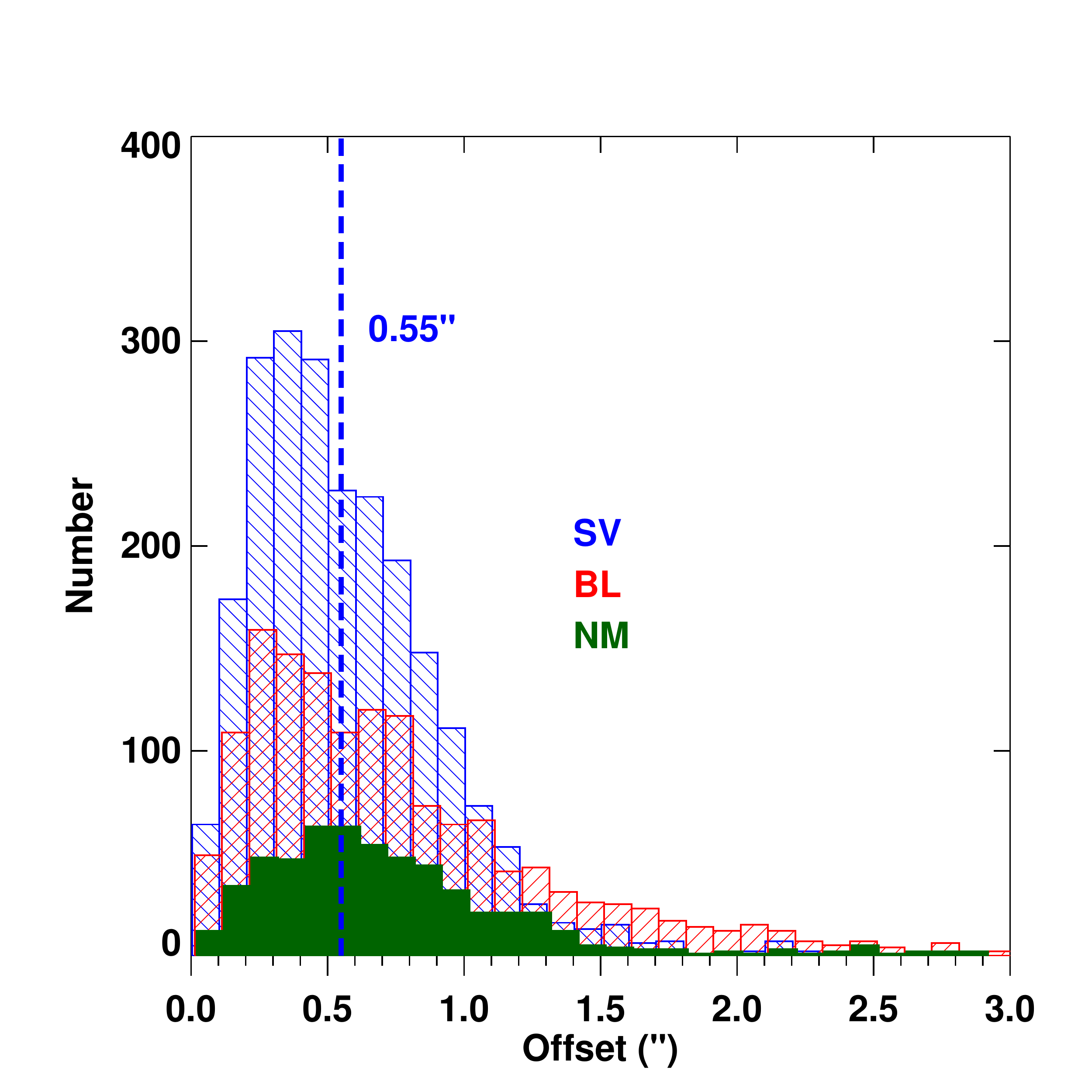}
\caption{Host galaxy offset distributions for SV, BL, and NM sources in arcsec.  Dashed line indicates the offset above which a source is considered ``off-nuclear''.} \label{QualityoffsetAGN} 
\end{figure}

We combine the light-curve classifications, with host galaxy offsets, in order to define a robust photometrically selected sample of AGN and SNe, from nuclear SV and off-nuclear BL, respectively.  In order to determine our cut-off for off-nuclear sources, we first fit the offset distribution for all the sources, with a bimodal distribution (Fig.~\ref{offset}), to derive a nuclear (predominantly AGN) population with $\mu_{nuc}=0.26,\sigma_{nuc}=0.14$, and off-nuclear (predominantly SNe) population with $\mu_{off-nuc}=0.48,\sigma_{off-nuc}=0.37$.  SNe can be coincident with galaxy nuclei due to the limited spatial resolution of the images. AGN, however, should not have significant offsets from their host galaxy centers, unless of course, they are more exotic objects such as recoiling supermassive black holes, or dual AGN.  Therefore, objects with offsets greater than $\mu_{nuc}+2\sigma_{nuc}=0.54$ are most likely to be SNe. For offsets $<0.54"$ the AGN population is more than $97\%$ complete, and there is negligible contamination in the supernova population by AGN, beyond this offset. Following this line of reasoning, we use the AGN from our verification set to determine the nuclear offset distribution, shown in Fig.~\ref{AGNGALEXoffset}. This is fitted with a Gaussian and leads to a $\mu_{AGN}+2\sigma_{AGN}=0.55$ cut-off for AGN, which we adopt for demarcating nuclear sources from off-nuclear ones. The offset distribution for each variability class is shown in Fig.~\ref{QualityoffsetAGN}. The distribution of SV offsets is broader than that of the verification set AGN, however, the broader distribution likely reflects the larger errors in the image difference and host galaxy centroids for fainter AGN, that are not represented in the verification set.  The BL distribution is seen to extend well beyond the nuclear AGN distribution, as would be expected for SNe. There are also a significant number of SVs with offsets $>0.55"$, however, we find that these sources are faint with a  mean $g\sim21$, resulting in their offsets being poorly determined. Table \ref{Classifications} shows the number of sources in each variability class divided into nuclear (offset $<\mu_{\rm nuc}+2\sigma_{\rm nuc}=0.55\arcsec$) and off-nuclear (offset $> 0.55 \arcsec$). We designate the nuclear SV to be AGN, of which there are 1233, and the off-nuclear BL to be SNe, of which there are 812. In the next section, we use these variability/offset selected AGN and SNe, to define photometric priors for their easy identification in future surveys.

\section{Characterization of Variability Selected Extragalactic AGN and SNe} \label{Hosts}

For upcoming multi-band, multi-epoch surveys such as LSST, 
we have shown that light-curve characterization combined with
host galaxy offsets is a robust way to select AGN,
SNe, and other exotic events, and does not require data external to the 
survey such as spectroscopic follow-up. 
Using all the $g_{P1}, r_{P1}, i_{P1}, z_{P1}$ bands offers a redundancy
that increases the confidence of source
classification.  With our photometrically
selected samples of AGN and SNe, we now characterize their
key observed source and host galaxy properties,
with the hopes of finding priors that can accelerate their identification
in future surveys.

\begin{figure}[htp]
\centering
\includegraphics[scale=0.3]{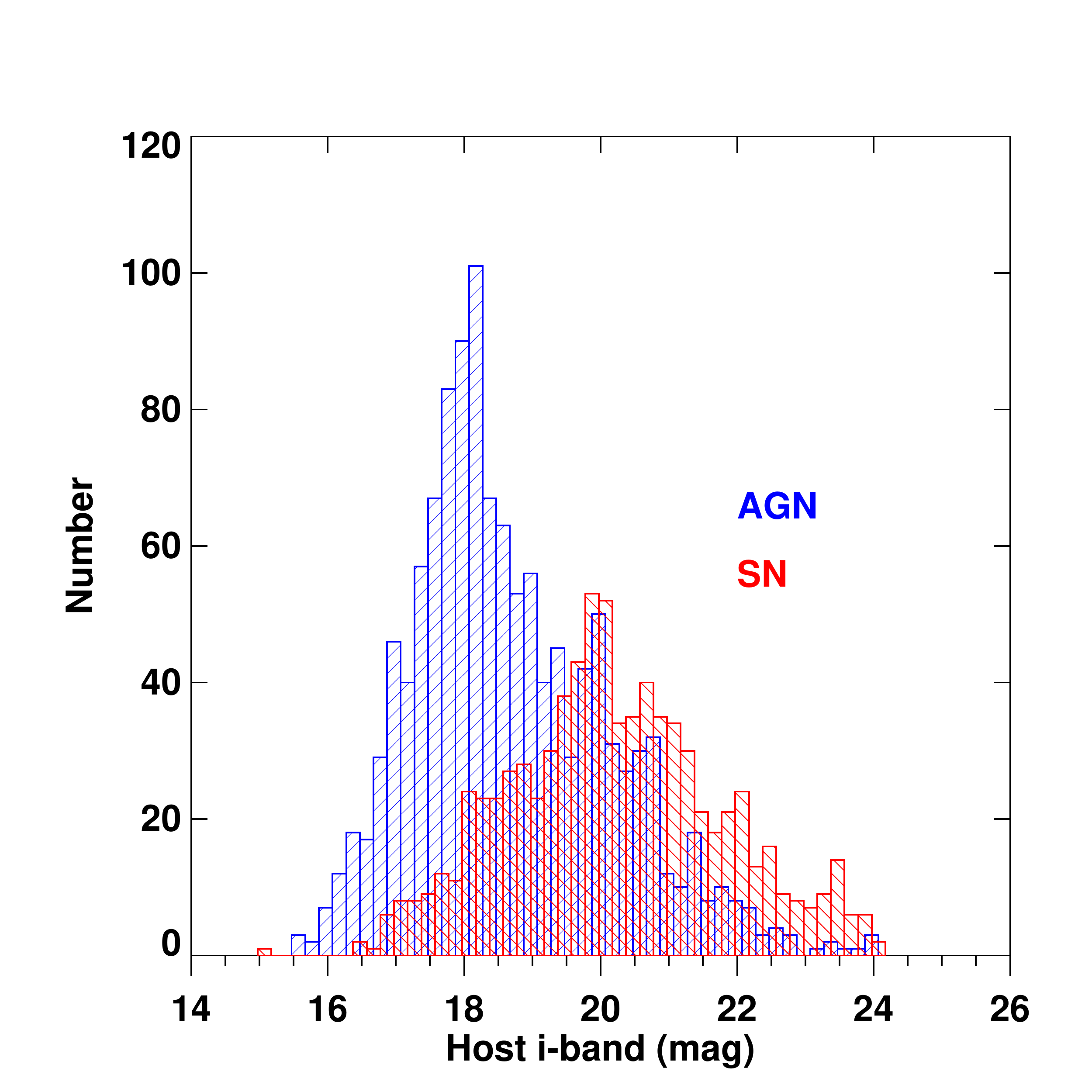}
\caption{Distribution of SN and AGN host galaxy i-band magnitudes. AGN host galaxies are $\approx3$ mag brighter than SN host galaxies.} \label{AGNI} 
\end{figure}

We use the $i_{P1}$-band to characterize the host galaxy magnitudes
of our sources, since the $i_{P1}$-band has the highest signal-to-noise
ratio amongst all the PS1 bands, and contamination of host galaxy flux
by a central AGN is minimized as compared to the bluer bands.
Fig.~\ref{AGNI} shows the distribution of host galaxy $i_{P1}$ for AGN and SN. AGN host galaxies appear
significantly brighter in the i-band than the SN host galaxies. Preliminary redshift estimates of the transient alert host galaxies indicate that SN host galaxies have a larger mean redshift distribution \citep{Heinis2014} as compared to the AGN host galaxies, thereby resulting in the observational bias.  

\begin{figure}[htp]
\centering
\includegraphics[scale=0.3]{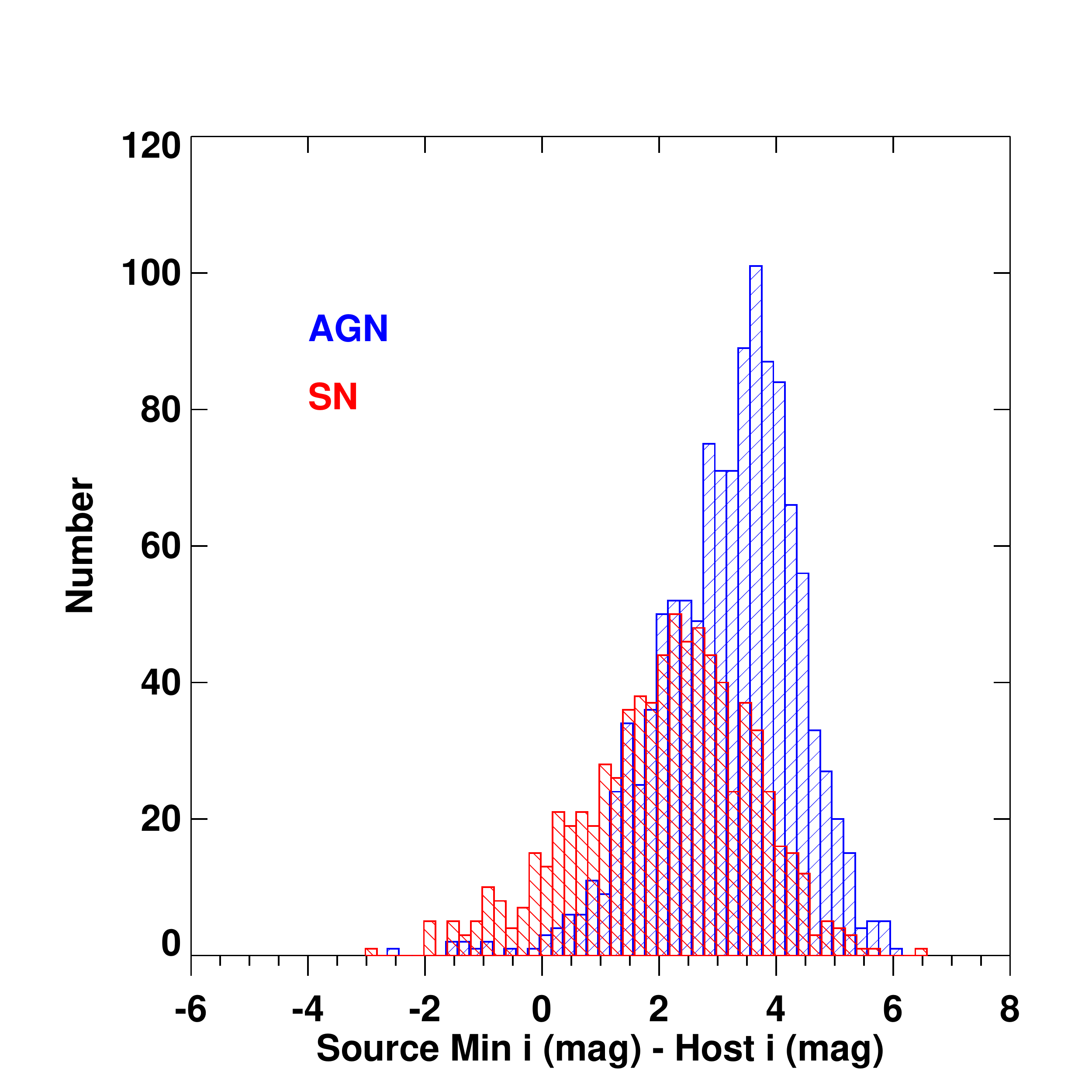}
\caption{Distribution of the differences between the minimum i-band difference magnitude and the host galaxy i-band magnitude for all source types. AGN fluxes are typically much fainter relative to their host galaxies with typical ($i_{\rm AGN}-i_{host})\approx4$ mag, while SNe are typically $3$ mag fainter than their host galaxies in the i-band. } \label{AGNhistminiI} 
\end{figure}

AGN themselves are much fainter in difference flux as compared to their host galaxy flux, and we can use this to further separate the AGN from the SN using the distribution of the differences between the minimum source i-band difference-magnitude and the host magnitude ($i_{min}-i_{host}$) (Fig.~\ref{AGNhistminiI}). AGN peak variability amplitudes are significantly fainter ($\approx4$ mag) relative to their host galaxies, as compared to that for SNe ($\approx2$ mag), consequently being more difficult to detect. 

\begin{figure}[htp]
\centering
\includegraphics[scale=0.3]{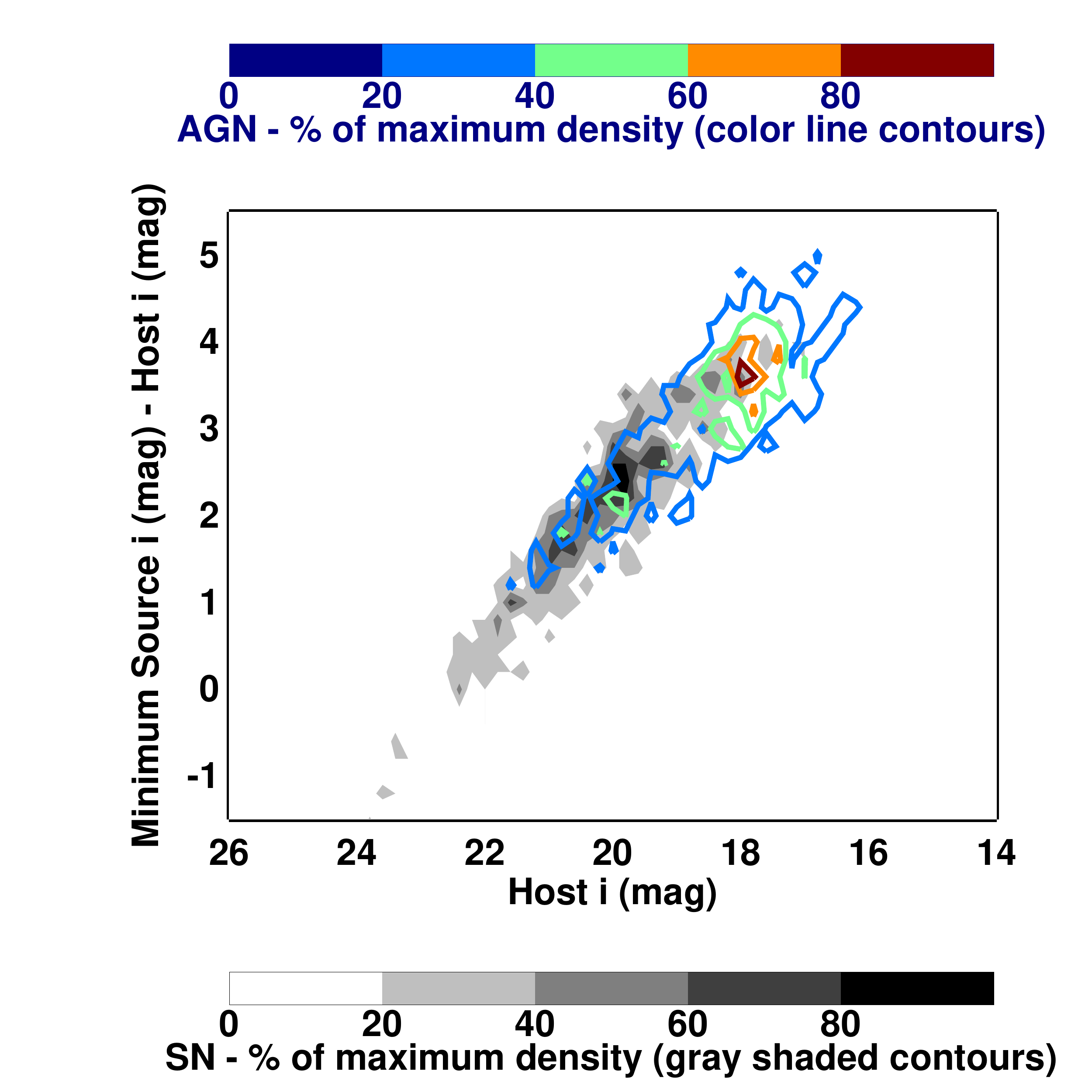}
\caption{Distribution of AGN (contours in blue to red) and SN (contours in gray to black) $i_{host}$ and $i_{min}-i_{host}$. A clear separation can be seen between the highest density regions of the two source types.} \label{AGNSNminiII} 
\end{figure}

We find that by using only $i_{\rm host}$ and ($i_{\rm min}-i_{\rm host}$), we can compute informative priors for the source-types from their relative probabilities of occurrence. Fig.\ref{AGNSNminiII} shows the contours of AGN and SN in $i_{host}$ and $i_{min}-i_{host}$ space. Although the AGN and SN distributions overlap in this space, there is a clear divide between their highest density regions, making it possible to separate them and assign relative probabilities in the overlap regions. Approximating and smoothing the SNe and AGN, $i_{\rm host}$ and ($i_{\rm min}-i_{\rm host}$) distributions (in Fig.\ref{AGNI} and Fig.\ref{AGNhistminiI} respectively) by Gamma distributions, we obtain their respective joint probability distributions in both parameters as: 

\begin{align} 
p_{AGN} &= \gamma(i_{min}-i_{host},k=25.360,\theta=0.230) \nonumber\\
&\times \gamma(i_{host},k=4.911,\theta=0.651)\\
p_{SN} &= \gamma(i_{min}-i_{host},k=12.852,\theta=0.400) \nonumber\\
&\times \gamma(i_{host},k=11.080,\theta=0.469)
\end{align}

\begin{figure*}[htp]
\centering
\includegraphics[width=7in,height=4.5in]{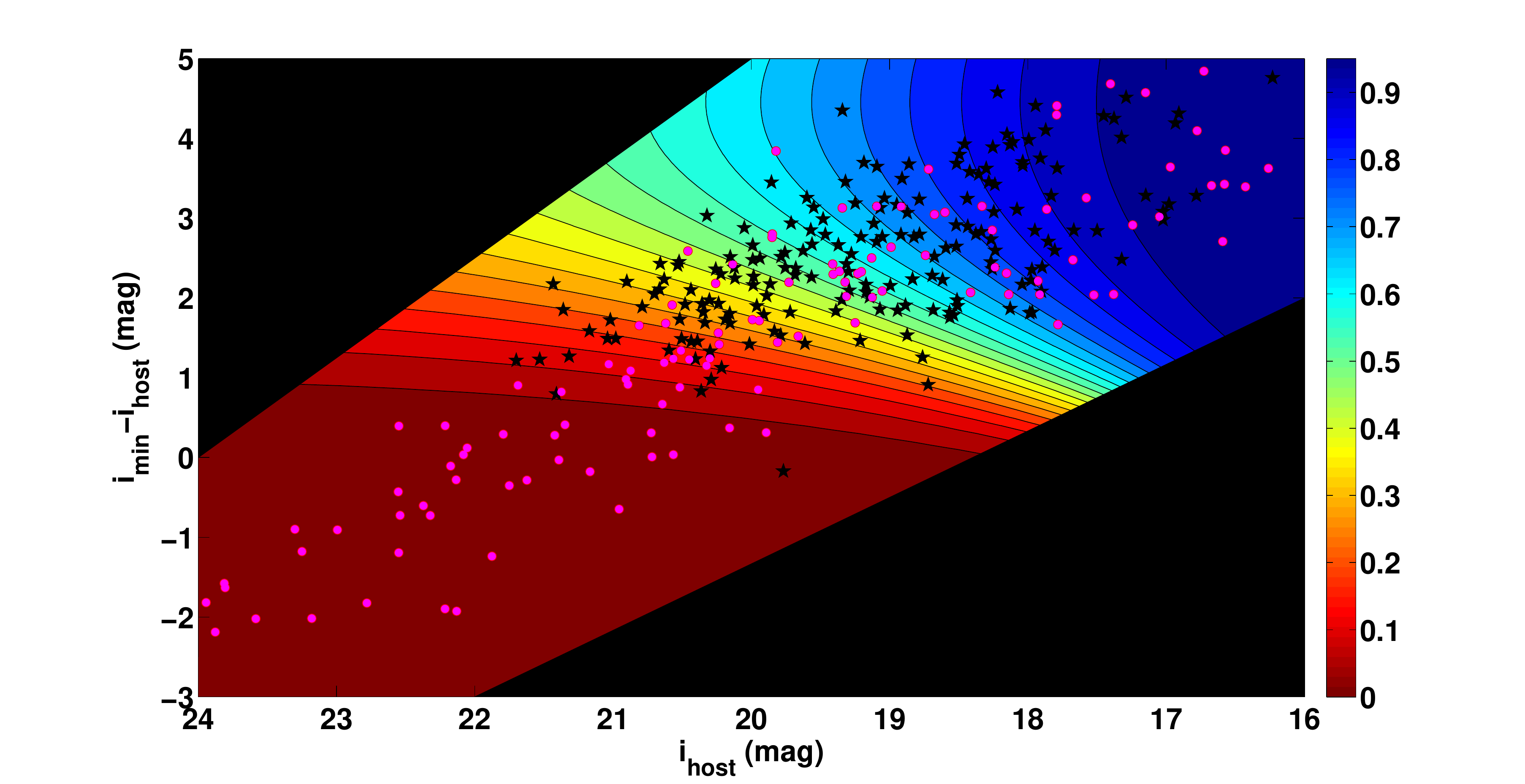}
\caption{Smoothed distribution of relative AGN probability (Eq.\ref{relAGN}) in the $i_{host}$ - $i_{min}-i_{host}$ plane. The probability distributions, derived from the density of the photometrically selected AGN and SN samples in each parameter, are smoothed and approximated by Gamma distributions. The overall distribution is obtained by multiplying the distributions in each parameter. The verification-set AGN (black stars) and SNe (magenta circles) are plotted for reference.} \label{AGNSNprob} 
\end{figure*}

If $N_{AGN}$ and $N_{SN}$ are the observed number of AGN and SN, the relative AGN likelihood for any set of values $i_{host}$ and $i_{min}-i_{host}$ is given by 

\begin{eqnarray}  
p_{\rm AGN | AGN,SN)} = \frac{N_{AGN}p_{AGN}}{N_{AGN}p_{AGN}+N_{SN}p_{SN}} \label{relAGN}
\end{eqnarray}

where $p_{\rm AGN | AGN,SN)}$ is the probability that an object is an AGN, given that it is either an AGN or an SN. Assuming that the number of AGN and SNe scale linearly with the number of SV and BL sources respectively, we obtain $N_{AGN}=2262$ and $N_{SN}=1529$.  Fig.\ref{AGNSNprob} is a smoothed version of Fig.\ref{AGNSNminiII} and shows the contours of $p_{\rm AGN | AGN,SN}$. SNe at their brightest, being brighter than AGN, and also being the source-type that dominate alerts from fainter galaxies, typically occupy smaller $i_{min}-i_{host}$ and larger $i_{host}$ (redder contours). Whereas, AGN fluxes being smaller compared to their host galaxy fluxes, and from less distant galaxies, occupy larger $i_{min}-i_{host}$ and smaller $i_{host}$ (bluer contours). The probability of a SN at any given point in this parameter space is $1-p_{\rm AGN | AGN,SN}$. The verification set AGN (black stars) and SNe (magenta circles) are plotted for reference. 

For the problem of classification in real-time from a large data stream such as the LSST transient alerts, we have found that for a magnitude-limited survey, simply using the i-band peak source magnitude and i-band host magnitude as priors, one can produce a robust preliminary AGN vs. SN classification, in order to help filter out a sample for more tedious methods such as spectroscopic or time-series identification of sources.    

\section{Conclusions} \label{Conclusions}

In this paper, we discussed a multi-band difference-flux time-series based method for the classification of $4361$ PS1 MD extragalactic difference-image sources into stochastic and bursting. Using a star-galaxy catalog to select extragalactic sources, we classify them into SV and BL sources using band-wise difference-flux characterization. Although this method can use actual or difference-magnitude time-series, difference-flux time-series are preferred over difference magnitudes which are log scaled, thus circumventing the problem of negative difference-flux excursions in SV light curves, for which magnitudes cannot be defined. We use multiple BL models to model the shapes of BL light curves, an OU process to model SV light curves, and a No-Model to identify white-noise dominated light curves. Since the models only attempt to differentiate between coherent single-burst type behavior and stochastic variability, they do not assume any underlying physical processes for the sources, making the method widely applicable. The use of multiple BL models is justified for statistical redundancy in the parameterizations of the light curves, as well as for modeling the gamut of shapes of BL light curves. We find that using all 3 BL models in the place of any one or two of them, improves the purity and completeness of the variability classifications of our AGN and SN verification sets. We estimate the model fitnesses using their estimated corrected-Akaike information criteria, and their leave-out-one-cross-validation likelihoods in each filter. The use of these independent derived statistical measures, one of which is suited to simply assess light curve shape characteristics, and the other to assess the overall robustness of the model, works to fortify the derived classifications. 

We then construct decision vectors $RV_{i,f}$ for each source, based on the AICc and the logarithm of the LOOCV for all the time-series models, which are combined in two clustering steps across the sources, and classified using a supervised K-means clustering method to arrive at the final filter-wise classifications; we filter out the NM sources in the first step, and then we separate out the SV and BL sources in the second. The use of time-series in multiple bands increases the reliability of our classifications. We then define two quality measures $C_i$ and $D_i$, which are filter-wise averages of the final clustering classification parameters, in which space the SV and BL can be separated. We find that our method results in $183$ verification set AGN being classified with $95.00\%$ purity and $57.92\%$ completeness, and $130$ verification set SNe classified with $90.97\%$ purity and $93.89\%$ completeness. We use our method to classify all the extragalactic difference-detection alerts into $2262$ SV,  $1529$ BL, and $570$ NM best-fit sources. We then construct a robust photometrically selected sample of $812$ SNe and $1233$ AGN, using a combination of light-curve class and host galaxy offset, to characterize their variability properties and the properties of their host galaxies, in order to better inform their real-time identification in photometric surveys. We find that simply the combination of i-band host magnitude and the i-band source magnitude can be used as robust preliminary indicators of source type. 

We demonstrate that our method can be used to separate SV from BL using the self-contained data (multi-epoch image-differencing and deep stacks) available in multi-band time domain surveys, such as PS1 and LSST. However, one could go further and use other variability based parameters in conjunction with our time-series method, together with host galaxy offsets, colors, and morphology, and external information from multi-wavelength catalog associations, in a larger, comprehensive hierarchical classification scheme to improve classification accuracy, characterize known sub-classes of sources, as well discover new classes of sources. In addition to the classification of variables and transients into broad general classes and particular sub-classes via the use of exact models, ensemble studies of their general properties can be readily performed; for example, the general properties of the host galaxies of AGN and SNe; the rates and properties of SNe and their subclasses; the variability timescales and amplitudes of AGN and their subclasses, and subsequently, the estimation of the black hole mass function; are some of the questions that can be readily answered using the model-fit parameter distributions for the respective classes. In the era of wide-field synoptic surveys generating millions of transient alerts per night, such self-contained photometric identification,  classification, and characterization of transients based on light-curve characteristics and host galaxy properties will be essential.  

\acknowledgments

We thank the referee and the scientific editor for their insightful comments and suggestions. The Pan-STARRS1 Surveys (PS1) have been made possible through contributions of the Institute for Astronomy, the University of Hawaii, the Pan-STARRS Project Office, the Max-Planck Society and its participating institutes, the Max Planck Institute for Astronomy, Heidelberg and the Max Planck Institute for Extraterrestrial Physics, Garching, The Johns Hopkins University, Durham University, the University of Edinburgh, Queen's University Belfast, the Harvard-Smithsonian Center for Astrophysics, the Las Cumbres Observatory Global Telescope Network Incorporated, the National Central University of Taiwan, the Space Telescope Science Institute, the National Aeronautics and Space Administration under Grant No. NNX08AR22G issued through the Planetary Science Division of the NASA Science Mission Directorate,  the National Science Foundation under Grant No. AST-1238877, and the University of Maryland.

\bibliographystyle{apj}
\bibliography{msfinalpdf}

\end{document}